\documentclass[a4paper,11pt]{article}
\pdfoutput=1 

\usepackage{jcappub} 

\usepackage[T1]{fontenc} 
\usepackage{lineno}

\usepackage{amsmath}
\usepackage{relsize}
\usepackage{commath}
\usepackage{mleftright}
\usepackage{bm}
\usepackage{tabularx}

\usepackage{comment}
\usepackage{ulem}

\usepackage{subcaption}
\usepackage{wrapfig}
\usepackage{graphicx} 
\graphicspath{{figures/}} 
\usepackage{cprotect}

\usepackage{array}
\usepackage{booktabs}
\usepackage{multirow}
\usepackage{threeparttable}

\newcommand{\ensemble}[1]{{\widetilde{#1}}}
\newcommand{\ellmax}{\ell_\mathrm{max}}
\newcommand{\ellmin}{\ell_\mathrm{min}}
\newcommand{\Cov}{\mathrm{Cov}}

\newcommand{\myvector}[1]{\bar{#1}}
\newcommand{\mymatrix}[1]{\mathbf{#1}}

\newcommand{\myset}[1]{{#1}}
\newcommand{\rev}[1]{{}}

\newcommand{\LB}{\textit{LiteBIRD}}

\newcommand{\LiteBIRD}{\LB}

\newcommand{\PySM}{\texttt{PySM}}
\newcommand{\nilc}{{\tt NILC}}
\newcommand{\NILC}{\nilc}
\newcommand{\bsecret}{{\tt B-SeCRET}}

\makeatletter
\newcommand\notsotiny{\@setfontsize\notsotiny\@vipt\@viipt}
\makeatother

\begin{document}

\title{In-flight polarization angle calibration for LiteBIRD: blind challenge and cosmological implications}

\author[1,2,3]{N.\,Krachmalnicoff,}
\author[4]{T.\,Matsumura,}
\author[5,6]{E.\,de\,la\,Hoz,}
\author[7]{S.\,Basak,}
\author[8,9]{A.\,Gruppuso,}
\author[10]{Y.\,Minami,}
\author[1,2,3]{C.\,Baccigalupi,}
\author[11]{E.\,Komatsu,}
\author[5]{E.\,Martínez-González,}
\author[5]{P.\,Vielva,}
\author[12]{J.\,Aumont,}
\author[13]{R.\,Aurlien,}
\author[14,4]{S.\,Azzoni,}
\author[12]{A.\,J.\,Banday,}
\author[5]{R.\,B.\,Barreiro,}
\author[15,16]{N.\,Bartolo,}
\author[17,18]{M.\,Bersanelli,}
\author[19]{E.\,Calabrese,}
\author[20,21]{A.\,Carones,}
\author[5]{F.\,J.\,Casas,}
\author[22,23,24]{K.\,Cheung,}
\author[25,4]{Y.\,Chinone,}
\author[26,27]{F.\,Columbro,}
\author[26,27]{P.\,de\,Bernardis,}
\author[5,6]{P.\,Diego-Palazuelos,}
\author[28]{J.\,Errard,}
\author[8,9]{F.\,Finelli,}
\author[13]{U.\,Fuskeland,}
\author[13]{M.\,Galloway,}
\author[29,30]{R.\,T.\,Genova-Santos,}
\author[31]{M.\,Gerbino,}
\author[14,4]{T.\,Ghigna,}
\author[32]{S.\,Giardiello,}
\author[13]{E.\,Gjerløw,}
\author[33,34,4,35]{M.\,Hazumi,}
\author[36]{S.\,Henrot-Versillé,}
\author[24]{T.\,Kisner,}
\author[26,27]{L.\,Lamagna,}
\author[31]{M.\,Lattanzi,}
\author[37]{F.\,Levrier,}
\author[38]{G.\,Luzzi,}
\author[17]{D.\,Maino,}
\author[26,27]{S.\,Masi,}
\author[20,21]{M.\,Migliaccio,}
\author[12]{L.\,Montier,}
\author[8]{G.\,Morgante,}
\author[12]{B.\,Mot,}
\author[34]{R.\,Nagata,}
\author[39,40]{F.\,Nati,}
\author[32,31]{P.\,Natoli,}
\author[32,31,41]{L.\,Pagano,}
\author[26,27]{A.\,Paiella,}
\author[8,9]{D.\,Paoletti,}
\author[28]{G.\,Patanchon,}
\author[26,27]{F.\,Piacentini,}
\author[38]{G.\,Polenta,}
\author[39,40]{D.\,Poletti,}
\author[20,24]{G.\,Puglisi,}
\author[42]{M.\,Remazeilles,}
\author[29,30]{J.\,Rubino-Martin,}
\author[43]{M.\,Sasaki,}
\author[44]{M.\,Shiraishi,}
\author[45]{G.\,Signorelli,}
\author[46,4]{S.\,Stever,}
\author[45,47]{A.\,Tartari,}
\author[36]{M.\,Tristram,}
\author[44]{M.\,Tsuji,}
\author[12]{L.\,Vacher,}
\author[13]{I.\,K.\,Wehus,}
\author[39,40]{and M.\,Zannoni}
\author[ ]{\\LiteBIRD Collaboration.}
\affiliation[1]{International School for Advanced Studies (SISSA), Via Bonomea 265, 34136, Trieste, Italy}
\affiliation[2]{INFN Sezione di Trieste, via Valerio 2, 34127 Trieste, Italy}
\affiliation[3]{IFPU, Via Beirut, 2, 34151 Grignano, Trieste, Italy}
\affiliation[4]{Kavli Institute for the Physics and Mathematics of the Universe (Kavli IPMU, WPI), UTIAS, The University of Tokyo, Kashiwa, Chiba 277-8583, Japan}
\affiliation[5]{Instituto de Fisica de Cantabria (IFCA, CSIC-UC), Avenida los Castros SN, 39005, Santander, Spain}
\affiliation[6]{Dpto. de Física Moderna, Universidad de Cantabria, Avda. los Castros s/n, E-39005 Santander, Spain}
\affiliation[7]{School of Physics, Indian Institute of Science Education and Research Thiruvananthapuram, Maruthamala PO, Vithura, Thiruvananthapuram 695551, Kerala, India}
\affiliation[8]{INAF - OAS Bologna, via Piero Gobetti, 93/3, 40129 Bologna (Italy)}
\affiliation[9]{INFN Sezione di Bologna, Viale C. Berti Pichat, 6/2 – 40127 Bologna Italy}
\affiliation[10]{Research Center for Nuclear Physics, Osaka University, Ibaraki, Osaka, 567-0047, Japan}
\affiliation[11]{Max-Planck-Institute for Astrophysics, Garching}
\affiliation[12]{IRAP, Universit$\acute{\rm e}$ de Toulouse, CNRS, CNES, UPS, (Toulouse), France}
\affiliation[13]{Institute of Theoretical Astrophysics, University of Oslo, Blindern, Oslo, Norway}
\affiliation[14]{Department of Physics, University of Oxford, Denys Wilkinson Building, Keble Road, Oxford OX1 3RH, United Kingdom}
\affiliation[15]{Dipartimento di Fisica e Astronomia “G. Galilei”, Universita` degli Studi di Padova, via Marzolo 8, I-35131 Padova, Italy}
\affiliation[16]{INFN Sezione di Padova, via Marzolo 8, I-35131, Padova, Italy}
\affiliation[17]{Dipartimento di Fisica, Universita' degli Studi di Milano, Via Celoria 16 - 20133, Milano, Italy}
\affiliation[18]{INFN Sezione di Milano, Via Celoria 16 - 20133, Milano, Italy}
\affiliation[19]{Cardiff University, School of Physics and Astronomy, Cardiff CF10 3XQ, UK}
\affiliation[20]{Dipartimento di Fisica, Universit\`{a} di Roma Tor Vergata, Via della Ricerca Scientifica, 1, 00133, Roma, Italy}
\affiliation[21]{INFN Sezione di Roma2, Universit\`{a} di Roma Tor Vergata, via della Ricerca Scientifica, 1, 00133 Roma, Italy}
\affiliation[22]{University of California, Berkeley, Department of Physics, Berkeley, CA 94720, USA}
\affiliation[23]{University of California, Berkeley, Space Science Laboratory,  Berkeley, CA 94720, USA}
\affiliation[24]{Lawrence Berkeley National Laboratory (LBNL), Computational Cosmology Center, Berkeley, CA 94720, USA}
\affiliation[25]{University of Tokyo, School of Science, Research Center for the Early Universe, RESCEU}
\affiliation[26]{Dipartimento di Fisica, Universit\`{a} La Sapienza, P. le A. Moro 2, Roma, Italy}
\affiliation[27]{INFN Sezione di Roma, P.le A. Moro 2, 00185 Roma, Italy}
\affiliation[28]{AstroParticle and Cosmology (APC) - University Paris Diderot, CNRS/IN2P3, CEA/Irfu, Obs de Paris, Sorbonne Paris Cit\'e, France}
\affiliation[29]{Instituto de Astrofísica de Canarias, E-38200 La Laguna, Tenerife, Canary Islands, Spain}
\affiliation[30]{Departamento de Astrofísica, Universidad de La Laguna (ULL), E-38206, La Laguna, Tenerife, Spain}
\affiliation[31]{INFN Sezione di Ferrara, Via Saragat 1, 44122 Ferrara, Italy}
\affiliation[32]{Dipartimento di Fisica e Scienze della Terra, Universit\`a di Ferrara, Via Saragat 1, 44122 Ferrara, Italy}
\affiliation[33]{High Energy Accelerator Research Organization (KEK), Tsukuba, Ibaraki 305-0801, Japan}
\affiliation[34]{Japan Aerospace Exploration Agency (JAXA), Institute of Space and Astronautical Science (ISAS), Sagamihara, Kanagawa 252-5210, Japan}
\affiliation[35]{The Graduate University for Advanced Studies (SOKENDAI), Miura District, Kanagawa 240-0115, Hayama, Japan}
\affiliation[36]{Universit\'e Paris-Saclay, CNRS/IN2P3, IJCLab, 91405 Orsay, France}
\affiliation[37]{Laboratoire de Physique de l’$\acute{\rm E}$cole Normale Sup$\acute{\rm e}$rieure, ENS, Universit$\acute{\rm e}$ PSL, CNRS, Sorbonne Universit$\acute{\rm e}$, Universit$\acute{\rm e}$ de Paris, 75005 Paris, France}
\affiliation[38]{Space Science Data Center, Italian Space Agency, via del Politecnico, 00133, Roma, Italy}
\affiliation[39]{University of Milano Bicocca, Physics Department, p.zza della Scienza, 3, 20126 Milan Italy}
\affiliation[40]{INFN Sezione Milano Bicocca, Piazza della Scienza, 3, 20126 Milano, Italy}
\affiliation[41]{Université Paris-Saclay, CNRS, Institut d’Astrophysique Spatiale, 91405, Orsay, France.}
\affiliation[42]{University of Manchester, Manchester M13 9PL, United Kingdom}
\affiliation[43]{Remeis-Observatory and Erlangen Centre for Astroparticle Physics, FAU Erlangen-N\"urnberg, Sternwartstr. 7, 96049 Bamberg, Germany}
\affiliation[44]{National Institute of Technology, Kagawa College}
\affiliation[45]{INFN Sezione di Pisa, Largo Bruno Pontecorvo 3, 56127 Pisa (Italy)}
\affiliation[46]{Okayama University, Department of Physics, Okayama 700-8530, Japan}
\affiliation[47]{Dipartimento di Fisica, Università di Pisa, Largo B. Pontecorvo 3, 56127 Pisa}

\emailAdd{nkrach@sissa.it}
\emailAdd{tomotake.matsumura@ipmu.jp}

\abstract{We present a demonstration of the in-flight polarization angle calibration for the JAXA/ISAS second strategic large class mission, \LiteBIRD, and estimate its impact on the measurement of the tensor-to-scalar ratio parameter, $r$, using simulated data.
We generate a set of simulated sky maps with CMB and polarized foreground emission, and inject instrumental noise and polarization angle offsets to the 22 (partially overlapping) \LiteBIRD\ frequency channels. Our in-flight angle calibration relies on nulling the $EB$ cross correlation of the polarized signal in each channel. This calibration step has been carried out by two independent groups with a blind analysis, allowing an accuracy of the order of a few arc-minutes to be reached on the estimate of the angle offsets. Both the corrected and uncorrected multi-frequency maps are propagated through the foreground cleaning step, with the goal of computing clean CMB maps. We employ two component separation algorithms, the Bayesian-Separation of Components and Residuals Estimate Tool (\bsecret), and the Needlet Internal Linear Combination (\NILC). We find that the recovered CMB maps obtained with algorithms that do not make any assumptions about the foreground properties, such as \NILC, are only mildly affected by the angle miscalibration. However, polarization angle offsets strongly bias results obtained with the parametric fitting method. Once the miscalibration angles are corrected by $EB$ nulling prior to the component separation, both component separation algorithms result in an unbiased estimation of the $r$ parameter. While this work is motivated by the conceptual design study for \LiteBIRD, its framework can be broadly applied to any CMB polarization experiment. In particular, the combination of simulation plus blind analysis provides a robust forecast by taking into account not only detector sensitivity but also systematic effects. }

\maketitle
\flushbottom

\section{Introduction}
\label{sec:intro}

The measurement of temperature and polarization anisotropies in the cosmic microwave background (CMB) plays a crucial role in modern cosmology \cite{komatsu/etal:2014,PlanckCosmo:2020,adachi/etal:2020,aiola/etal:2020,sayre/etal:2020,dutcher/etal:2021,BICEP:2021xfz}.
In recent years, the \textit{Planck} satellite  has observed the CMB signal over the entire celestial sphere in both total intensity and polarization, returning a picture of our Universe in  excellent agreement with the standard cosmological constant ($\Lambda$) + Cold Dark Matter (CDM) cosmological model ~\cite{Planck2018Overview}.

The focus has now shifted primarily to the measurement of the imprint of primordial gravitational waves \cite{grishchuk:1975,starobinsky:1979} predicted by the inflationary paradigm on the CMB polarized signal \cite{seljak/zaldarriaga:1997,kamionkowski/kosowsky/stebbins:1997}. The theory of cosmic inflation, which assumes a period of accelerated expansion in the very early evolution of the Universe, was originally proposed to explain unresolved problems in cosmology \cite{Guth,sato,linde:1982,albrecht/steinhardt:1982}; it also predicts that, if primordial perturbations were generated from vacuum fluctuations in the early Universe, their wavelength would be stretched to macroscopic length scales by an exponential expansion phase \cite{Mukhanov:1981xt,Hawking:1982cz,Starobinsky:1982ee,Guth:1982ec,Bardeen:1983qw}, thus leaving an imprint on the CMB signal. In particular, tensor perturbations in the metric (gravitational waves) would generate a curl component in the CMB polarized signal, called $B$ modes, at angular scales larger than about $1^{\circ}$. There has been no detection of this faint primordial $B$-mode signal yet. Its amplitude is parameterized by the  tensor-to-scalar ratio, $r$,  the amplitude of which is directly related to the energy scale of inflation \cite{Marc}. The current upper limit, $r<0.036$~(95\%~C.L.), 
has been obtained from the combination of the \textit{Planck} and BICEP/Keck Array data~\cite{tristram2020planck,BICEP:2021xfz}. This 
corresponds to a $B$-mode signal of amplitude $\sim 50$ nK.

The ongoing effort in the CMB community to reach the instrumental sensitivity needed to probe the primordial $B$-mode signal is driving significant advancement in focal-plane technology. 
Current operational ground-based experiments, including SPTpol, the advanced ACTpol, BICEP/Keck, and POLARBEAR, employ focal planes of about $10^3$ superconducting detector arrays~\cite{sptpol,Thornton_2016,10.1117/12.2233894,Kaneko2020}.
The next generation of ground-based projects, such as the Simons Observatory, the South Pole Observatory, and eventually CMB Stage IV (CMB-S4) experiment, will employ a total of $\sim10^4$ and $>10^5$ detectors, respectively~\cite{SO_Science_Ade_2019,10.1117/12.2561995,s4collaboration2020cmbs4}. 
In space, \LiteBIRD \cite{hazumi/etal:2019}, the second strategic large-class mission
selected by the Institute of Space and Astronautical Science (ISAS)/Japan Aerospace Exploration Agency (JAXA), is scheduled to observe the sky from the second Lagrangian (L2) point of the Earth-Sun system, in the late 2020s with $\sim5000$ detectors. 
These experiments will achieve a sensitivity on the order of a few to several $\mu$K$\cdot$arcmin with multi-frequency coverage, reaching a statistical noise level comparable to, or lower than, the $B$-mode signal of the weak gravitational lensing effect on the CMB $E$-mode polarization \cite{zaldarriaga/seljak:1998}. Such experiments could  enable the first ever detection of the signature of primordial gravitational waves with $r\gtrsim10^{-3}$ in the CMB.

Pushing instrumental sensitivities to these levels, by installing complex focal planes on telescopes, leads to new challenges in the control of instrumental systematic effects.
In particular, one of the major possible systematics is related to the need for accurate calibration of the intrinsic polarization angles of the detectors.
The Stokes parameters of linear polarization can be written as $Q\pm iU=P\exp(\pm 2i\gamma)$, where $P$ and $\gamma$ are the polarization intensity and angle, respectively. If the polarization angle of a detector has an uncalibrated offset $\alpha$, the observed angle, $\gamma^{\rm o}$, would shift from the true value $\gamma$ to an incorrect one, $\gamma^{\rm o}=\gamma+\alpha$.  The observed Stokes parameters would then be related to the true ones by $Q^{\rm o}\pm iU^{\rm o}=(Q\pm iU)\exp(\pm 2i\alpha)$.

Following Refs. \cite{zaldarriaga/seljak:1997,kamionkowski/kosowsky/stebbins:1997b}, we use spin-2 spherical harmonics to expand the Stokes parameters in the $\hat{n}$ direction as $Q(\hat{n})\pm i U(\hat{n})=-\sum_{\ell m} (E_{\ell m} \pm i B_{\ell m})\,_{\pm2}Y_{\ell m}(\hat{n})$. It then follows  that the observed $E$- and $B$-mode spherical harmonic coefficients are given by $E_{\ell m}^{\rm o}\pm iB_{\ell m}^{\rm o}=(E_{\ell m}\pm iB_{\ell m})\exp(\pm 2i\alpha)$, or
\begin{eqnarray}
\label{eq:Elm}
E_{\ell m}^{\rm o}&=& E_{\ell m}\cos(2\alpha)-B_{\ell m}\sin(2\alpha)\,,\\
\label{eq:Blm}
B_{\ell m}^{\rm o}&=& E_{\ell m}\sin(2\alpha)+B_{\ell m}\cos(2\alpha)\,.
\end{eqnarray}
Thus, any uncertainty associated with the polarization angle with respect to the sky coordinates leads to a mixing of $E$- and $B$-mode polarization signals \cite{shimon/etal:2008,miller/shimon/keating:2009}. For example, even if the true sky contained no $B$-mode signal, we would observe a spurious $B$-mode power spectrum of $C_\ell^{BB,{\rm o}}=\sin^2(2\alpha)C_\ell^{EE}$.
This leakage of the bright $E$-mode to the much weaker $B$-mode signal is a major source of systematic uncertainty, and can introduce a possible bias in the tensor-to-scalar ratio measurement \cite{PhysRevD.67.043004,WMAP9year}, since the required accuracy in the knowledge of the detector polarization angle to achieve a sensitivity of $r=10^{-3}$ can be as demanding as a few arcminutes~\cite{litebird_polarangle_requirement}. 

Miscalibration of the instrumental polarization angle is not the only instrumental systematic effect that can cause mixing of polarization modes. A similar effect can arise, for example, from the presence of non-ideality in the optical beam shape \cite{shimon/etal:2008,miller/shimon/keating:2009}, which therefore should also be known to high accuracy. In this paper, however, we focus on the implementation of a strategy to mitigate the impact of the former effect. The conventional approach for calibrating the detector polarization angle has been to employ an external polarized source, e.g., a polarizing grid with a known polarization orientation with respect to the polarimeters.
Such a device is placed in either the near or far field from a telescope, and is used to calibrate the polarization angles prior to or during the observing campaign. However, this strategy can be employed only if the required accuracy is at the level of $1^{\circ}$ degree \cite{Takahashi_2010}, unless we can substantially improve upon the precision of the current generation of calibrators. Detailed modeling of the optics system can also help to improve knowledge of the instrumental polarization \cite{koopman:2018,choi/etal:2020}.
Another possibility to calibrate the absolute polarization angle is to use a polarized sky source, e.g., Tau A and the Galactic diffuse emission~\cite{Matsumura_2010,Aumont2020_Crab,Masi:2021jju}. 
The current measurement accuracy of these sources, however, is not sufficiently high to allow the calibration of the absolute polarization angle at the targeted sensitivity for the tensor-to-scalar ratio. 
While we do not exclude these options when observational data from future ground-based and balloon-borne CMB telescopes is available, we do not rely on them in this paper. 

An alternative approach to calibrating the absolute polarization angle with sub-degree accuracy is to null the $EB$ cross-correlation, $C^{EB}_{\ell}$. This approach can be used either under the assumption that no cosmological $EB$ signal is present~\cite{Keating_2012}, or generalized to the case where intrinsic $EB$ correlation exists in the sky signal~\cite{minami/etal:2019,minami:2020,minami/komatsu:2020a}. In this paper we carry out analysis under the first assumption. Even if there were no intrinsic $EB$ correlation (which is the case in the standard model of cosmology), the miscalibration angle would yield a spurious $EB$ power spectrum given by
\begin{equation}
\label{eq:EB}
C_\ell^{EB,{\rm o}}=\frac12(C_\ell^{EE}-C_\ell^{BB})\sin(4\alpha)\,,
\end{equation}
as derived from Eqs.~(\ref{eq:Elm}) and (\ref{eq:Blm}).
Thus, we can use $C_\ell^{EB,{\rm o}}$ to solve for $\alpha$ given the prior knowledge of the intrinsic $C_\ell^{EE}-C_\ell^{BB}$, with an accuracy limited by cosmic variance and the noise level of the experiments. This approach, called ``self-calibration'', has been applied to 
BICEP1, BICEP2/Keck, POLARBEAR, and SPTpol data sets~\cite{PhysRevD.89.062006,PhysRevLett.112.241101,P_A_R_Ade_2014,bianchini/etal:2020}. In this paper, we further study the feasibility of polarization angle calibration by this method. 
Due to the stringent requirement on the uncertainty in the knowledge of $\alpha$ for future experiments such as \LiteBIRD\ and CMB-S4, establishing a reliable calibration method will solidify the overall calibration strategy and could potentially reduce the required accuracy at the hardware level. Although in this work we assumed non-zero 

To increase the reliability of our study, we carry out the analysis in a blind fashion.  Moreover, in this paper we restrict ourselves to the $EB$ field as a tracer of the spurious correlation induced by the rotation angle. In principle, one could also consider the $TB$ channel as complementary, though not totally independent, information. in this way we are conservative, by considering the main source of information, postponing the full exploitation of other sets of data to a future work.
In all previous work using the $EB$ self-calibration technique, the polarized CMB was assumed to be the dominant sky signal, and the effect of polarized foreground emission was ignored \cite{PhysRevD.89.062006,PhysRevLett.112.241101,P_A_R_Ade_2014,bianchini/etal:2020}.
In the absence of foregrounds, the technique is straightforward to implement. As demonstrated in Ref.~\cite{minami/etal:2019}, it is possible to extend the $EB$ self-calibration method in the foreground-dominated channel (see section~\ref{sec:method} for details). However, it is important to check whether this successful demonstration was a special case or if the method would be successful in general. It is particularly important is to avoid tuning the details of the method to mitigate the impact of foreground emission in the simulations. To avoid this potential human bias, we have decided to carry out our analysis in a blind fashion.
Specifically, two independent groups use independently developed tools to analyze the simulated data, without knowing the values of the offset angles injected into them.
We then ``open the box'' and compare the results with the true answer only after the completion of both independent analyses.

\begin{figure}[t]
    \centering
    \includegraphics[width=1\textwidth]{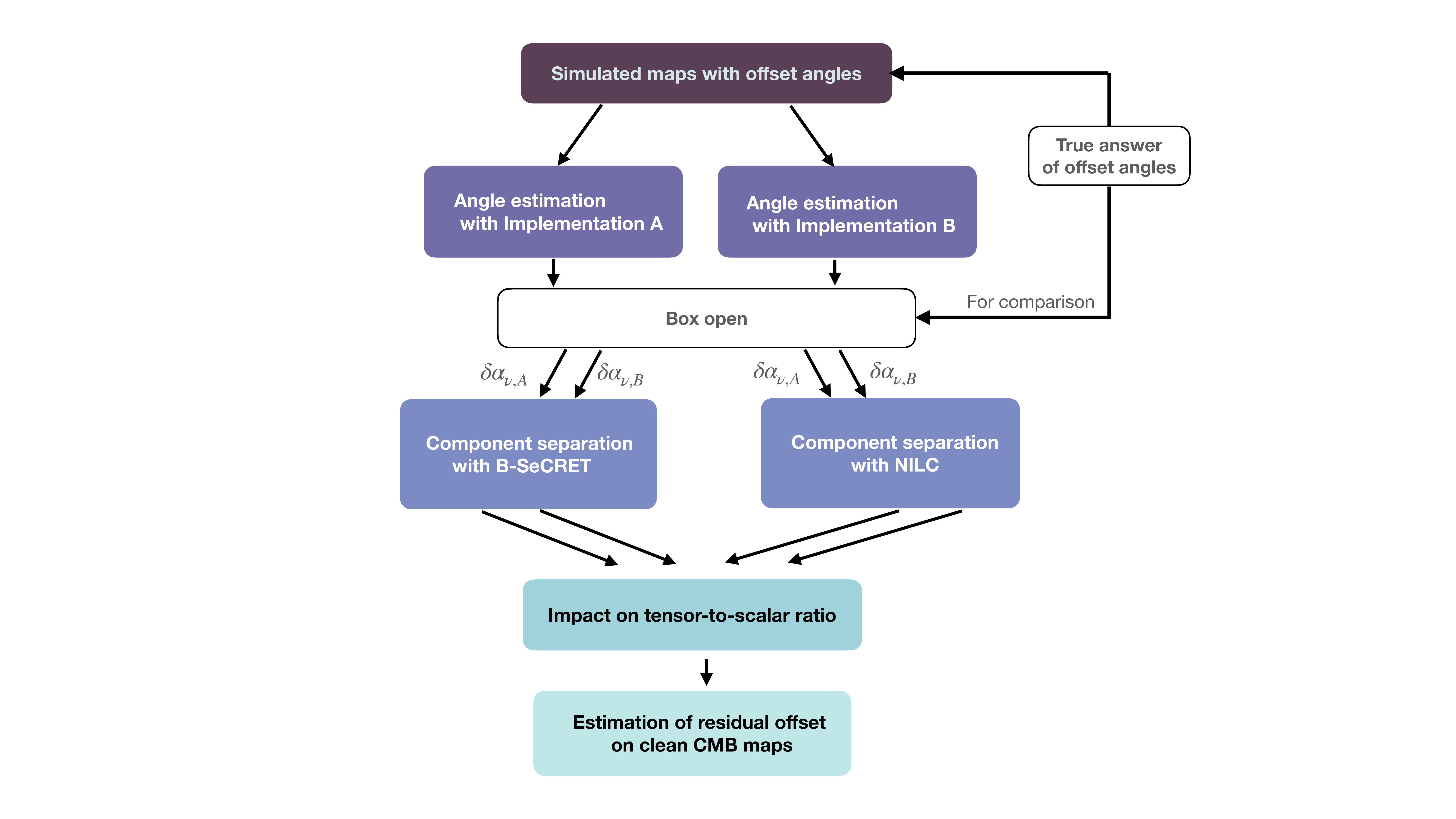}
    \caption{Block diagram of the data analysis steps presented in this paper.}
    \label{fig:blockdiagram}
\end{figure}

We also propagate the residual errors after calibration in the analysis pipeline through to the determination of $r$. 
While this exercise is applicable to any CMB polarization experiment, \rev{independently of its instrumental specifications,} we specifically apply it to the \LiteBIRD\ satellite as a test case, \rev{given that, among the forthcoming CMB experiments, it has the highest forecast sensitivity on the $r$ parameter and the broadest frequency coverage}. Figure~\ref{fig:blockdiagram} shows a block diagram describing the steps in our analysis.

The remainder of the paper is organized as follows. 
In section~\ref{sec:sim}, we describe the preparation of the simulated maps including polarization angle offsets. In section~\ref{sec:method}, we summarize two implementations (Implementation A and B) of the self-calibration method for determining the offset angles, and compare the results.
In section~\ref{sec:cs}, we study the impact of the calibration errors on the estimate of $r$ by propagating them in the component separation analysis step. In section~\ref{sec:discussion}, we discuss the validity and limitations of the study framework and conclude. In appendix~\ref{App:pol_methods}, we provide the details of the two calibration algorithms and in appendix~\ref{app:component_separation_methodologies}, we describe our component separation methods (\NILC\ and \bsecret).

\section{Simulated maps}
\label{sec:sim}
We assess the feasibility of $EB$-based in-flight polarization angle calibration on a set of simulated multi-frequency maps. 
These maps include the sky emission, from both the CMB and Galactic foreground signals, instrumental noise, and the injected polarization angle offsets. 
In this study we use the instrumental specifications of the \LiteBIRD\ satellite, as reported in table~\ref{tab:LB_instrument}~\cite{Sugai2020}. \par

\begin{table}
      \caption{Instrumental specification used to produce the simulations in this work.}
         \label{tab:LB_instrument}
     $$ 
       \begin{threeparttable}
         \begin{tabular}{lccc} 
            \hline\hline
            \noalign{\smallskip}\noalign{\smallskip}
        Channel & Frequency & FWHM  & Pol. sensitivity\\  
        name & [GHz] & [arcmin] & $\mu$K-arcmin\\
            \noalign{\smallskip}
             \hline
            \noalign{\smallskip}
              LFT-40 & 40 & 69.3 & 59.29 \\
              LFT-50 & 50 &  56.8 & 32.78 \\
              LFT-60 & 60 & 49.0 & 25.76 \\
              LFT-68a & 68 & 41.6 & 21.60 \\
              LFT-68b & 68 & 44.5 & 23.53 \\
              LFT-78a & 78 & 36.9 & 18.59 \\
              LFT-78b & 78 & 40.0 & 18.45 \\ 
              LFT-89a & 89 & 33.0 & 16.95 \\
              LFT-89b & 89 & 36.7 & 15.03 \\
              LFT-100 & 100 & 30.2 & 12.93 \\
              LFT-119 & 119 & 26.3 & 9.79 \\
              LFT-140 & 140 & 23.7 & 9.55 \\
              \noalign{\smallskip}
             \hline
            \noalign{\smallskip}
              MFT-100 & 100 & 37.8 & 9.67 \\
              MFT-119 & 119 & 33.6 & 6.41 \\
              MFT-140 & 140 & 30.8 & 7.02 \\
              MFT-166 & 166 & 28.9 & 5.81 \\
              MFT-195 & 195 & 28.0 & 7.12 \\
               \noalign{\smallskip}
             \hline
            \noalign{\smallskip}
            HFT-195 & 195 & 28.6 & 15.66 \\
            HFT-235 & 235 & 24.7 & 15.16 \\
            HFT-280 & 280 & 22.5 & 17.98 \\
            HFT-337 & 337 & 20.9 & 24.99 \\
            HFT-402 & 402 & 17.9 & 49.90 \\
            \hline
             \noalign{\smallskip}
            \hline
         \end{tabular}
     \end{threeparttable}
     $$ 
      \end{table}

\subsection{Sky models}
We generate the CMB signal as a set of Gaussian realizations of the \textit{Planck} 2018 best-fit $\Lambda$CDM model without tensor modes ($r=0$) \citep{PlanckCosmo:2020}. The polarized Galactic foreground maps are generated through the Python Sky Model package (\PySM) \citep{2017MNRAS.469.2821T}. In particular, we consider the presence of polarized thermal dust and synchrotron emission. The polarized dust template implemented in \PySM, based on the \texttt{Commander} results from the \textit{Planck} 2015 data release \citep{planck2015-X}, was smoothed to an angular resolution of $2^{\circ}$, then Gaussian fluctuations added on smaller angular scales. The template at 353 GHz is scaled in frequency as a modified blackbody with spatially  uniform spectral parameters: $\beta_{\text{d}} = 1.54$ and $T_{\text{d}} = 20$ K (the so-called model \textit{d0} in \PySM). The synchrotron template corresponds to the \textit{WMAP} 9 year Stokes $Q$ and $U$ maps \citep{WMAP9year} at 23 GHz, smoothed to an angular resolution of $3^{\circ}$, to which Gaussian fluctuations were again added on small scales; the synchrotron spectral index is also spatially uniform with $\beta_{\text{s}}=-3$ (model \textit{s0} in \PySM). The templates adopted for both synchrotron and thermal dust emission do not show a detectable signal in the $EB$ correlation. We emphasize that, although updated templates exist for foreground emission, including information coming from new data \citep{Planck2018_IV} or developed with new algorithms \citep{Krachmalnicoff2021}, they have not been fully validated yet nor interfaced with the \PySM\ library. Therefore, since the main analysis of this work is not related to the evaluation of the optimal strategy for component separation, we simply retain the models already implemented in \PySM\ that are widely used by the community \citep{SO_Science_Ade_2019}.

Moreover, the assumption of spatially uniform spectral parameters is a simplification of the expected properties of the foreground emission. Recent observations at low and high frequencies have shown that both synchrotron and thermal dust emission show spatial variations in their spectral energy distributions (SEDs) for both polarization and total intensity, measured with varying levels of significance depending on the sky region \citep{planck2015-X, 2016A&A...596A.109P, 2018A&A...618A.166K}. However, we stress that, in this work, we focus on assessing the level of precision that can be reached in the estimation of polarization angle offsets through the nulling of the $EB$ correlation. The methods presented in section~\ref{sec:method} would not be affected by the spatial variation of the foreground SED, although the component separation step would be rendered significantly more complicated. We therefore choose to use the simplified sky model to separate the two problems. 
The simulations are produced for each of the 22 (partially overlapping) \LiteBIRD\ frequency channels, and the total signal (CMB plus foregrounds) is smoothed to the corresponding angular resolution (see table~\ref{tab:LB_instrument}). Noise is simulated as homogeneous Gaussian fluctuations corresponding to the sensitivity values reported in table~\ref{tab:LB_instrument}. No correlated noise is considered, this choice is made under the assumption of the use of a continuously rotating half-wave plate (HWP), which can up-covert the signal bandwidth above the low-frequency noise. The description of the polarimeter configuration is detailed in the next section. Note that we include the potential polarization angle miscalibration originating from the HWP. 
\rev{Moreover, since this work focuses specifically on the calibration of the polarization angle offset, we do not include in our simulations any other potential systematic effect.} We produce 10 different sets of simulations. In each set, we change the CMB and noise realization, as well as the polarization angle offsets, generated as described in the following section.

\subsection{Polarization angle offsets}
\begin{figure}[t]
    \centering
    \includegraphics[width=0.8\textwidth]{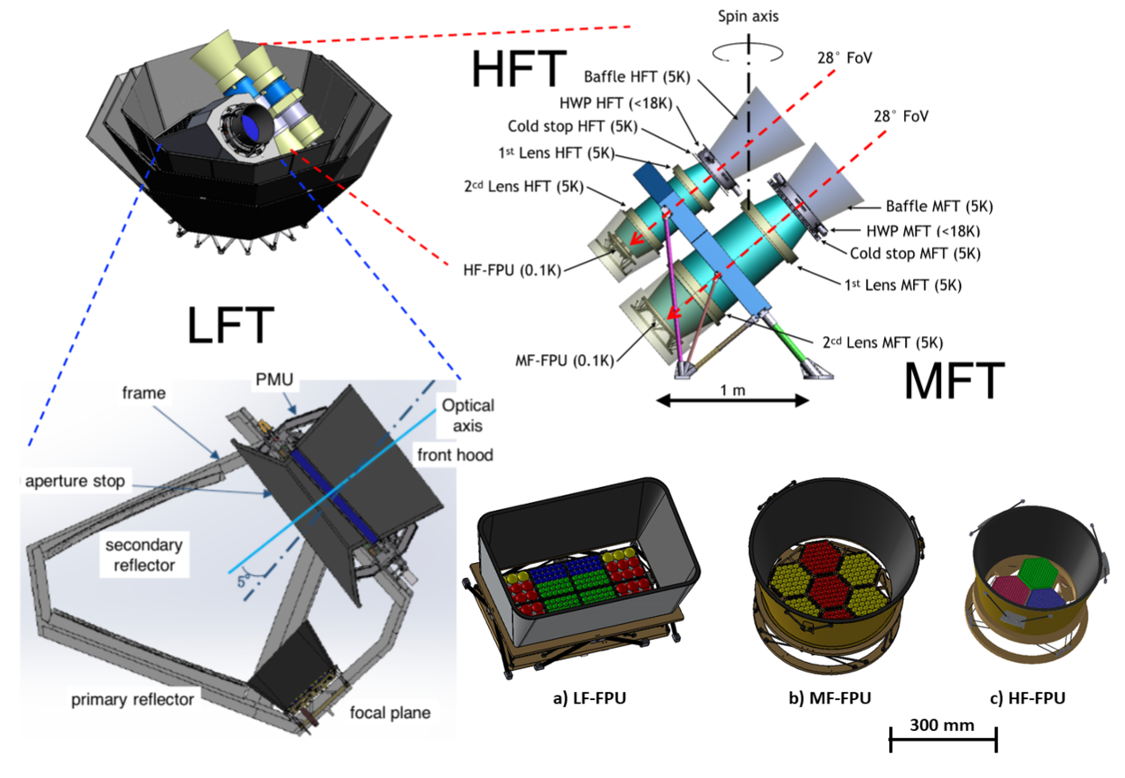}
    \caption{Overview of the \LiteBIRD\ satellite, the three telescopes (LFT, MFT, HFT), and the three focal planes for LFT (a), MFT (b), and HFT (c).}
    \label{fig:LBInstrumentoverview}
\end{figure}

As already mentioned, we use the \LiteBIRD\ instrument configuration as a test case for this study. 
\LiteBIRD\ consists of three telescopes -
the Low-Frequency Telescope (LFT), Mid-Frequency Telescope (MFT), and High-Frequency Telescope (HFT).
These three telescopes are cooled to a temperature of 5\,K. 
Structurally, LFT, MFT and HFT will be separately assembled, and each telescope will be integrated on the frame of a payload module that is supported by the rest of the satellite structure. 
Figure~\ref{fig:LBInstrumentoverview} shows an overview of the payload module. 
The overall mission concepts and configurations can be found in Refs.~\cite{10.1117/12.2563050}.

Each telescope consists of a continuously rotating HWP, an optical system (cross-Dragone for LFT, and two two-lens refractor systems for MFT and HFT), thermal filters, and a focal plane~\cite{10.1117/12.2561841,10.1117/12.2562243}.
Each focal plane contains an array of wafers and each wafer contains multiple beam-forming elements.
LFT and MFT employ a lenslet as the beam-forming element, with each lenslet containing six (three bands and two polarization states) transition-edge sensor (TES) bolometers. Similarly, HFT employs a feed-horn coupled antenna, each containing four TES bolometers (two bands and two polarization states) \cite{10.1117/12.2562978}.
As a result, the combination of the detectors at each band achieves the sensitivity listed in table~\ref{tab:LB_instrument}.

\begin{table}[t]
    \centering
    \begin{tabular}{ccc}
    \hline\hline
            \noalign{\smallskip}\noalign{\smallskip}
       Case  & Source & Range of injected offset angles \\
       \noalign{\smallskip}
       \hline
\noalign{\smallskip}
       1 & Global  & $ \le 5$~arcmin  \\
       2 & Telescope  & $ \le 1^{\circ}$ for each of LFT, MFT, and HFT, with no correlation \\
       3 & Observational band & $ \le 2^{\circ}$ for each 22-band unit with, no correlation  \\
 \hline
             \noalign{\smallskip}
            \hline
    \end{tabular}
    \caption{Summary of the ranges for absolute values of random offset angles. In all cases, we assume a uniform probability distribution within the stated range.}
    \label{tab:angleerrorcategory}
\end{table}

In the following, we describe potential sources of polarization angle offset in the \LiteBIRD\ polarimeters assumed in this study. 
We define an offset angle as the shift of the projected polarization-sensitive angle of the polarimeter on the sky from the designed orientation. 
We have considered three different sources of offset, as summarized in table~\ref{tab:angleerrorcategory}.
For Case 1, the simplest example is a global misalignment between the sky and satellite coordinates, which can be described by a single offset angle.
A candidate physical origin for this effect is the misalignment and/or miscalibration of a star tracker on the satellite frame with respect to the telescope mount. 
For Case 2, each of the LFT, MFT, and HFT can have its own independent offset angle, e.g., a misalignment of each telescope with respect to the common frame.
A potential physical origin for this is the uncertainty of the HWP position angle reconstruction in each telescope~\cite{10.1117/12.2576290,10.1117/12.2577818}.
For Case 3, we apply independent offset angles to the 22 observational frequency band units. 
Although there as 15 frequencies, the observational bands are divided into 22 units, so there are partial overlaps in frequency coverage between the three different telescopes. 
In addition, there are two types of
lenslet, with two different diameters, for the same observational frequency in LFT.
The  lenslet diameter changes the optical coupling, thus leading to different sensitivity. 
As a result, we treat them separately, labeled $a$ and $b$ in table~\ref{tab:LB_instrument}. 
Any wobbling effect caused by the sinuous antenna and a frequency-dependent polarization sensitive axis from the achromatic HWP can yield a rotational angle offset~\cite{2013PhDT.......591S,10.1117/1.JATIS.5.4.044008,Bao_2012}. 
Additionally, ignorance of the polarization angle rotation induced by the optical system itself can rotate the polarization angle over the focal plane by a few degrees~\cite{Kashima:18}. 
The focal plane position and the observational frequency are coupled, and thus this can also be a source of an offset angle due to miscalibration. 

We assume that all of these effects will be either calibrated or modeled with some imperfections at the component level prior to the observations, thus leading to the presence of offsets.
For each case, we draw offset angles randomly from a uniform probability distribution in the ranges reported in table~\ref{tab:angleerrorcategory}, and for each channel we sum the three contributions to obtain the final angle. 
We generate 10 different sets of miscalibrated maps,  simulated by directly rotating the Stokes $Q$ and $U$ maps in pixel space. 

While we limit the scope of our study to these three sources of miscalibration in this paper, 
we are aware that more cases could be considered in a more  realistic polarimeter model.
For example, sinuous antenna and achromatic HWPs have a frequency-dependent polarization angle effect~\cite{2013PhDT.......591S,10.1117/1.JATIS.5.4.044008,Bao_2012}, \rev{whose value could change within the frequency bandwidth of a given channel. This would add further complexity to both the simulation of the \LiteBIRD\ instrument, and the treatment of systematic effects, and will be addressed in future studies.}
Precision construction of polarimeters on a wafer allows us to accurately know the relative angles within one wafer in the same observational band. Therefore, we assume that the relative angles within a given wafer are well calibrated prior to the final calibration analysis using the $EB$ self-calibration method.

\section{Polarization angle calibration}
\label{sec:method}
In previous work~\cite{PhysRevD.89.062006,PhysRevLett.112.241101,P_A_R_Ade_2014,bianchini/etal:2020},
eq.~(\ref{eq:EB}) was used for calibrating the polarization angles with prior knowledge of $C_\ell^{EE}-C_\ell^{BB}\simeq C_\ell^{EE}$ of the CMB signal given by the best-fitting cosmological model. However, eq.~(\ref{eq:EB}) can be rewritten using only the {\it observed} difference of $E$- and $B$-mode power spectra, $C_\ell^{EE,{\rm o}}-C_\ell^{BB,{\rm o}}$ \cite{zhao/etal:2015,Gruppuso:2016nhj,minami/etal:2019}. We use this formulation in this paper.\par

While we assume throughout this paper that there is no $EB$ correlation in either the primordial or foreground signals, a new algorithm has been developed to  constrain the non-zero intrinsic $C^{EB}_{\ell}$ and the instrumental polarization angle offsets simultaneously~\cite{minami/etal:2019,minami:2020,minami/komatsu:2020a}. 
 Specifically, if we add prior knowledge of the CMB power spectra, we can determine $\alpha$ and an intrinsic $EB$ signal, perhaps due to the ``cosmic birefringence'' effect \cite{carroll:1998}, simultaneously \cite{minami/etal:2019}.
Our analysis can be simply extended to this case by replacing eq.~(\ref{eq:EBobs}) below with Eq.~(9) in Ref.  \cite{minami/etal:2019}, but here we prefer
to focus only on the determination of $\alpha$. For the simultaneous determination of $\alpha$ and the cosmic birefringence, as well as a discussion on the effect of the intrinsic foreground $EB$ correlation, see Refs.~\cite{2009PhRvD..80d3522P, minami/komatsu:2020b,abitbol/etal:2021,clark/etal:2021}.

Using the relationships between the observed $E$- and $B$-mode polarization and the intrinsic values, 
$E_{\ell m}^{\rm o}= E_{\ell m}\cos(2\alpha)-B_{\ell m}\sin(2\alpha)$ (eq.~(\ref{eq:Elm})) and $B_{\ell m}^{\rm o}= E_{\ell m}\sin(2\alpha)+B_{\ell m}\cos(2\alpha)$ (eq.~(\ref{eq:Blm})), the observed $E$- and $B$-mode power spectra are related to the intrinsic ones as
\begin{eqnarray}
C_\ell^{EE,{\rm o}} &=&
\cos^2(2\alpha)C_\ell^{EE}+\sin^2(2\alpha)C_\ell^{BB}\,,\\
C_\ell^{BB,{\rm o}} &=&
\sin^2(2\alpha)C_\ell^{EE}+\cos^2(2\alpha)C_\ell^{BB}\,,
\end{eqnarray}
in the absence of any physical $EB$ correlation.
Combining these with eq.~(\ref{eq:EB}), we obtain \cite{zhao/etal:2015,Gruppuso:2016nhj,minami/etal:2019}\footnote{Also see Ref. \cite{PhysRevLett.102.161302} for earlier work, which used the observed $EE$ and $BB$ power spectra but with $\sin(4\alpha)$ instead of $\tan(4\alpha)$, which would be the same in the limit $|\alpha|\ll 1$.}
\begin{equation}
\label{eq:EBobs}
C_\ell^{EB,{\rm o}}=
    \frac12(C_\ell^{EE,{\rm o}}-C_\ell^{BB,{\rm o}})\tan(4\alpha)\,.
\end{equation}
This expression no longer requires any prior knowledge of the intrinsic $E$- and $B$-mode power spectra, but the observed difference between them can be used to solve for $\alpha$. This allows us to retrieve the angle $\alpha$ not only in the presence of the CMB power spectra, which are  known accurately, but also foreground emission or any other sky components for which the theoretical power spectra are not available. \par

In this paper, we explore two different ways of implementing the estimation of $\alpha$ using eq.~(\ref{eq:EBobs}). We summarize them briefly in the following section, and provide details in appendix~\ref{App:pol_methods}. Our results are reported in section~\ref{Sec:angle_results}.

\subsection{Methods}
\label{Sec:methods}
We implemented the self-calibration technique in two ways (Implementations A and B) to estimate the polarization angle offsets in our simulated data. Both implementations use eq.~(\ref{eq:EBobs}) to solve for $\alpha$ by nulling the $EB$ cross correlation in the observed power spectra at different frequencies. In this section, we highlight the common assumptions of the two implementations as well as their differences. Detailed descriptions of the formalisms are given in appendix~\ref{App:pol_methods}. 

 Both implementations use a maximum likelihood approach, building the likelihood function from the observed power spectra, $C_{\ell}^{XY, {\rm o}}$, which are considered to follow a Gaussian distribution. In the case of Implementation A, only the auto-frequency power spectra of $BB$, $EE$ and $EB$ are considered, while in Implementation B all the possible frequency cross-spectra of the 22 channels (see table~\ref{tab:LB_instrument}) are included. Moreover, Implementation A makes the assumption of small angles for the $\alpha$ offsets, i.e., $\tan(\alpha) \to \alpha$. This approximation is not used in Implementation B. The covariance matrices are computed differently (see appendix~\ref{App:pol_methods}), with the assumption of being independent from the $\alpha$ parameter for Implementation A; in both cases the correlation between multipoles is neglected. The maximum of the likelihood function is found analytically and the uncertainties on $\alpha$ are estimated by the Fisher matrix for Implementation A, whereas for Implementation B the full posterior distribution is sampled with the publicly available Markov chain Monte Carlo (MCMC) algorithm \textsc{emcee}~\cite{foreman2013emcee}, with the marginalized $1\sigma$ uncertainties taken as errors on $\alpha$.\par

It is important to highlight that the two methods, corresponding to independent implementations by two different groups, also differ in details related to the computation of the power spectra, such as the multipole binning and the maximum $\ell$ value considered in the analysis. However, as already stressed, we carried out our analysis in a blind fashion, with the exact goal of avoiding the fine tuning of nuisance parameters in the implementations, in order to demonstrate that useful results can be achieved independently of those details.

\subsection{Results}
\label{Sec:angle_results}
\begin{figure}[t]
\centering
\includegraphics[width=\textwidth]{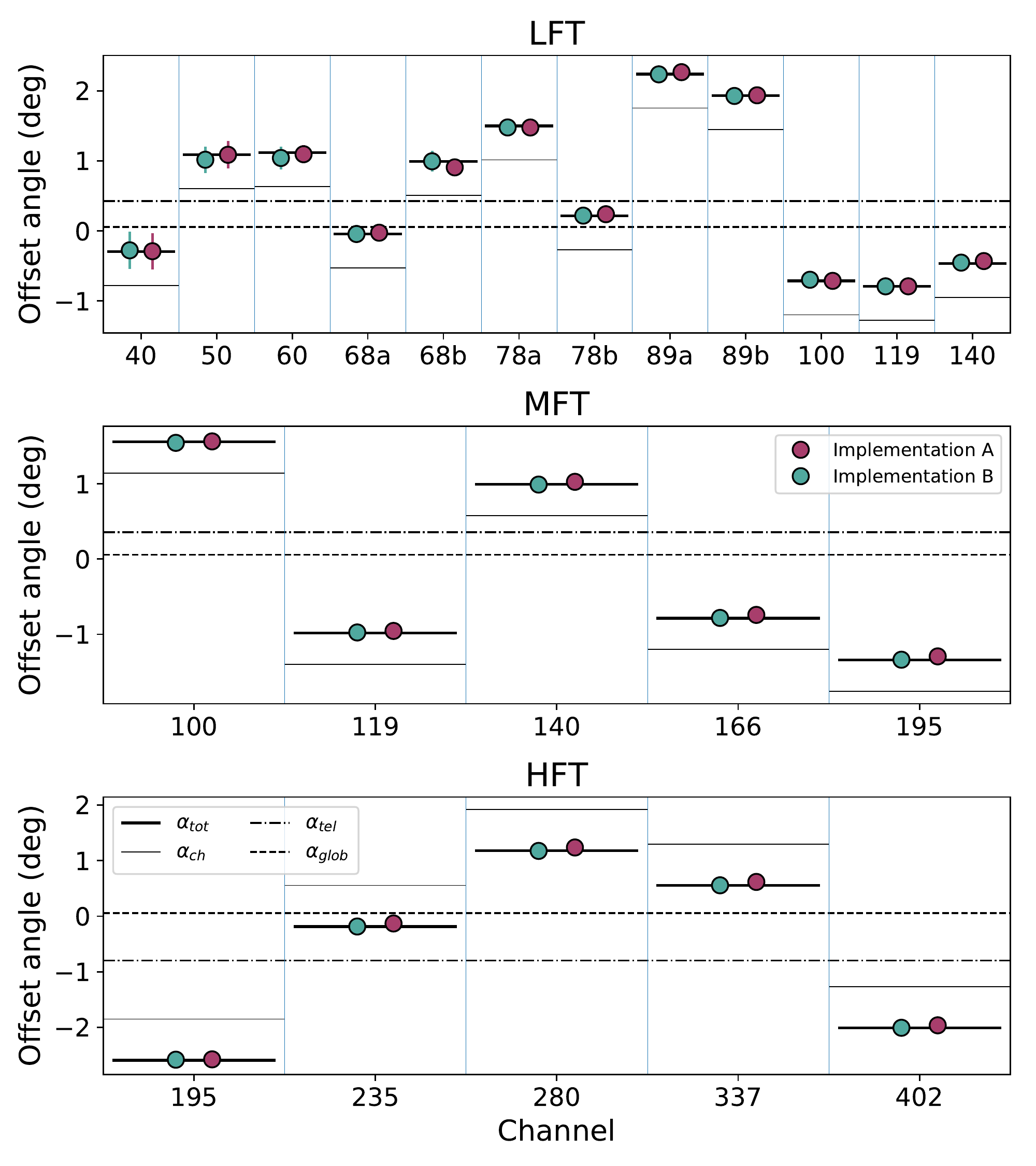}
\vspace{4 pt}
\caption{Comparison of the polarization angle estimates from Implementations A and B (filled circles) and the true injected offsets (black solid lines) for all the considered frequency channels and for one realization of the simulated maps. The error bars show the 3\,$\sigma$ uncertainties, which are smaller than the size of the filled circles except in the lowest frequency channels. The total true injected offset is obtained as the sum of the three possible sources considered: global misalignment (dashed lines), telescope offset (dashed-dotted lines) and wafer offsets (solid thin lines).}
\label{fig:comparison}
\end{figure}

\begin{figure}[t]
\centering
\includegraphics[width=13 cm]{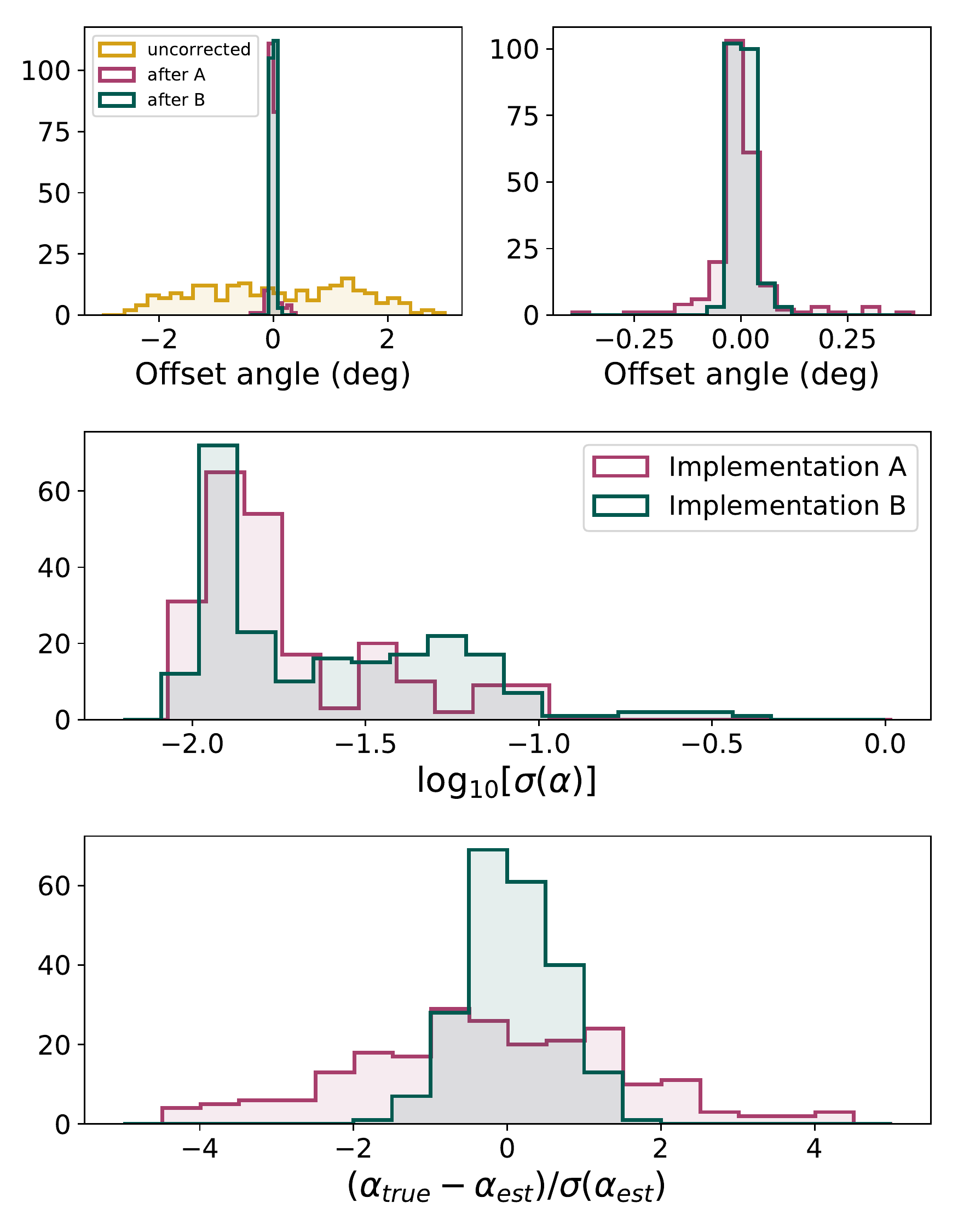}
\vspace{10pt}
\caption{Summary of the results for the estimation of the polarization angles with the two implementations presented in section~\ref{Sec:methods}. Histograms are computed from all the 10 different sets of simulated maps and the 22 frequency channels (i.e., a total of 220 realizations). The top panels show the distribution of the offset angles before (yellow histogram in the left panel) and after (pink and green histograms) corrections; the right figure shows a zoom on the relevant offset interval after correction. The middle panel shows the distribution of uncertainties in the estimated angles $\alpha_{\rm est}$. The bottom panel shows the distribution of the bias over the uncertainty of the estimation \rev{for each of the two implementations}.}
\label{fig:hist}
\end{figure}

We generated ten different sets of sky simulations, changing, in each of them, the noise and CMB realizations as well as the polarization angle offsets for each frequency channel. The values of the angle offsets were not revealed to the two analysis teams. Here we summarize the results of the angle calibration challenge.

In figure~\ref{fig:comparison},  we show the an example comparison of the true ($\alpha_{\rm true}$; black solid lines)  and estimated ($\alpha_{\rm est}$; filled circles) angles, for one of the ten realizations (results for the other cases are similar). We find that both implementations perform similarly, leading to estimated values for the offset angles close to the true ones for all frequency channels. The uncertainties are small (note that the figure reports $3\,\sigma$ uncertainties), with the error bars being visible only for the lowest frequency channels.

To better characterize and compare the results, figure~\ref{fig:hist} shows histograms of all of the angle realizations and frequency channels (for a total of 220 estimations).  The top panels show the distribution of the injected offsets ($\alpha_{\rm true}$; yellow histograms in the left panel) and those of the residuals, {\bf $\alpha_{\rm true}-\alpha_{\rm est}$}, after the self-calibration (coloured histograms in both panels). For Implementation A, the maximum absolute residual offset is at the level of $23'$, with 90\% of the cases below $6'$. For Implementation B, the absolute errors are smaller, lower than $2'$ in 90\% of the cases, and reach a maximum of  around $6'$. 

In the middle panel of figure~\ref{fig:hist}, we show the distributions of the estimated uncertainties on the reconstructed angles. We find that they are similar, which indicates that Implementation A may underestimate the uncertainties.  This is confirmed in the bottom panel, where the distributions of $\alpha_{\rm true}-\alpha_{\rm est}$ in units of the estimated uncertainties are shown. For Implementation B all the estimated angles are within $2\sigma$ of the true ones, while the distribution is broader for Implementation A, with some outliers at more than $4\sigma$.  During the post-processing following the unblinding of the challenge results, we found that the Fisher uncertainties used in Implementation A were underestimated due to the use of a non-optimal estimator of the ensemble average power spectra involved in the calculation of the likelihood's covariance matrix. This estimator did not properly model the foreground contribution, leading to an underestimation of the uncertainty especially at the lowest/highest frequency channels, where the foreground contribution is larger. \par
 In the following sections, we describe how we propagate the residual angle offsets into the data analysis pipeline and evaluate their impact on the  measurements of $r$.

\section{Component Separation}
\label{sec:cs}
The goal of our in-flight polarization angle calibration task is to reduce systematic errors in the determination of the tensor-to-scalar ratio, $r$. Here, we test the improvement on the measurements when Implementations A and B of the self-calibration technique are used to correct the miscalibration angles $\alpha$ in multi-frequency maps, and the residual offsets are propagated through the component separation step, with the goal of obtaining clean CMB maps. We apply component separation strategies to the following sets of maps, in order to characterize our results.
\begin{itemize}
    \item {\bf Non-rotated maps}: the original signal plus noise maps without any injected rotation of the polarization angle.
    \item {\bf Rotated maps}: the previous maps with the corresponding rotation offsets applied to the sky signal.
    \item \textbf{Derotated maps}: the maps after derotating the rotated maps with the solutions obtained from Implementations A and B. We derotate the sum of the signal and noise maps.
\end{itemize}

We adopt two complementary foreground-cleaning techniques that provide a complete framework for analyzing the propagation of rotation angle errors to the cosmological parameter estimation. The first is a map-based algorithm that adopts, and fits for, a parameterized model for the frequency dependence of the sky components.  The SED parameters and the amplitudes of the different sky signals are fitted in each resolution element by exploiting the multi-frequency observations. Several implementations of this procedure have been studied in the literature (see Ref. \cite{Planck2018_IV} and references therein), such as {\tt Commander}\footnote{\url{https://github.com/Cosmoglobe/Commander}} and {\tt FGBuster}\footnote{\url{https://fgbuster.github.io/fgbuster/index.html}}. In this paper, we use the procedure called \bsecret\ (Bayesian-Separation of Components and Residuals Estimate Tool) \cite{de2020detection}, described in detail in appendix~\ref{app:Bsecret}. 

The second component separation procedure, based on an internal linear combination (ILC) algorithm \cite{1996MNRAS.281.1297T,Bennett:2003ca,2003PhRvD..68l3523T}, does not assume any prior knowledge of the foreground SED, but rather minimizes the variance of a linear mixture of the multi-frequency data with the constraint of retaining the black body frequency scaling for the CMB.  In the implementation used in this work, called Needlet Internal Linear Combination (\NILC), the variance minimization is performed in both the spatial and harmonic domains. This allows us to take into account the non-uniform behavior of foregrounds, captured by the variation of the ILC coefficients over the sky and as a function of angular scale. In this work, we adopt the implementation used in the three data releases of the \textit{Planck} experiment \cite{2016A&A...594A...9P,Planck2018_IV}, as described in detail in appendix~\ref{sec:nilc_method}. 

In the following, we summarize the results of the analysis, on the sets of simulated maps described above, with \bsecret\ (section~\ref{sec:pf}) and \NILC\ (section~\ref{sec:nilc}). 
Before this, let us first discuss the expected impacts of the polarization angle miscalibration on the component separation process. When the foreground SED is uniform, we can approximate the component separation procedure as a weighted linear sum of the frequency channel maps, with coefficients that do not depend on sky location, but 
can depend on the multipole scale because of the different beams and noise levels of the frequency channels. 
We emphasize that this assumption is only adopted here to simplify the discussion: both \bsecret\ and \NILC\ allow for spatially-varying SEDs.
In the limit of small angles ($|\alpha_i|\ll 1$ for each $i$th channel), we write the spherical harmonic coefficients of the clean CMB $E$ and $B$ modes as
\begin{eqnarray}
    E_{\ell m}^{\rm clean}
    &=& \sum_{i=1}^{N_{\rm ch}}{\omega_{\ell,i}^E}E_{\ell m}^{{\rm o},i}\approx 
    E_{\ell m}^{\rm CMB} + 
    \sum_{i=1}^{N_{\rm ch}}{\omega_{\ell,i}^E}
    \left[E_{\ell m}^{{\rm FG},i}
    -2\alpha_i\left(B_{\ell m}^{\rm CMB}+B_{\ell m}^{{\rm FG},i}\right)
    \right]
    \,,\\
    B_{\ell m}^{\rm clean}
    &=& \sum_{i=1}^{N_{\rm ch}}{\omega_{\ell,i}^B}B_{\ell m}^{{\rm o},i}\approx 
    B_{\ell m}^{\rm CMB} + 
    \sum_{i=1}^{N_{\rm ch}}{\omega_{\ell,i}^B}\left[
    B_{\ell m}^{{\rm FG},i}+2\alpha_i\left(E_{\ell m}^{\rm CMB}+E_{\ell m}^{{\rm FG},i}\right)\right]\,,
\end{eqnarray}
where $N_{\rm ch}=22$ is the number of frequency channels and $\omega_{\ell,i}^X$ are the linear weights of the $i$th channel map of $X=(E,B)$ satisfying $\sum_i\omega_{\ell,i}^X=1$. Ignoring the intrinsic CMB $EB$ correlation and the noise bias, the ensemble average of the $EB$ and $BB$ power spectra of the clean map up to first order in $\alpha_i$ is given by
\begin{eqnarray}
C_\ell^{EB,{\rm clean}}
\nonumber
&=& 2\left(\alpha^B_{{\rm eff},\ell}C_\ell^{EE,{\rm CMB}}-\alpha^E_{{\rm eff},\ell}C_\ell^{BB,{\rm CMB}}\right)\\
\label{eq:cleanEB}
& &
+\sum_{ij}\omega_{\ell,i}^E\omega_{\ell,j}^B
\left[C_\ell^{E_iB_j,{\rm FG}}+2
\left(\alpha_jC_\ell^{E_iE_j,{\rm FG}}
-\alpha_iC_\ell^{B_iB_j,{\rm FG}}
\right)\right]\,,\\
\label{eq:cleanBB}
C_\ell^{BB,{\rm clean}}
&=& C_\ell^{BB,{\rm CMB}}
+\sum_{ij}\omega_{\ell,i}^B\omega_{\ell,j}^B
\left(C_\ell^{B_iB_j,{\rm FG}}+
4\alpha_iC_\ell^{E_iB_j,{\rm FG}}\right)\,,
\end{eqnarray}
where we have defined the new effective ($\ell$-dependent) angle, $\alpha_{{\rm eff},\ell}^X\equiv\sum_i\omega_{\ell,i}^X\alpha_i$.
We thus expect, even in the absence of the $EB$ correlation intrinsic to the foreground ($C_\ell^{E_iB_j,{\rm FG}}=0)$, that: 
\begin{itemize}
    \item[1.] The clean CMB map will have a non-zero $EB$ correlation, unless the angle miscalibration is corrected prior to the component separation;
    \item[2.] the $BB$ power spectrum of the clean CMB map will be affected by the angle miscalibration, if the presence of $\alpha_i$ affects the component separation (i.e., $\omega_{\ell,i}^X$). 
\end{itemize}
While the first is a trivial statement, the second is an interesting one. As we find below, the impact of $\alpha_i$ on the $BB$ power spectrum depends on the specific component separation method applied.

Moreover we note that the $\ell$ dependence of $\alpha_{{\rm eff},\ell}^X$ could potentially compromises the method decribed in~\cite{Sherwin:2021vgb} to cancel the angle miscalibration by comparing $EB$ from the reionization and recombination bumps.

\subsection{\bsecret}
\label{sec:pf}

We apply the \bsecret\ parametric component separation algorithm to the multi-frequency non-rotated, rotated and derotated maps, for each of  ten different sets of simulations. We fit a parametric model that includes seven parameters. Specifically, the thermal dust emission is parameterized with a modified blackbody SED, while the synchrotron radiation is parametrized with a power-law SED with spectral curvature. The details of the parametric model are reported in appendix~\ref{app:Bsecret}. Since the maps are simulated with spatially uniform SED parameters, we perform the fit on the whole sky jointly for $Q$ and $U$ maps, assuming that they share the same spectral parameters.\par

\begin{figure}[t]
   \centering
\includegraphics[width=\textwidth]{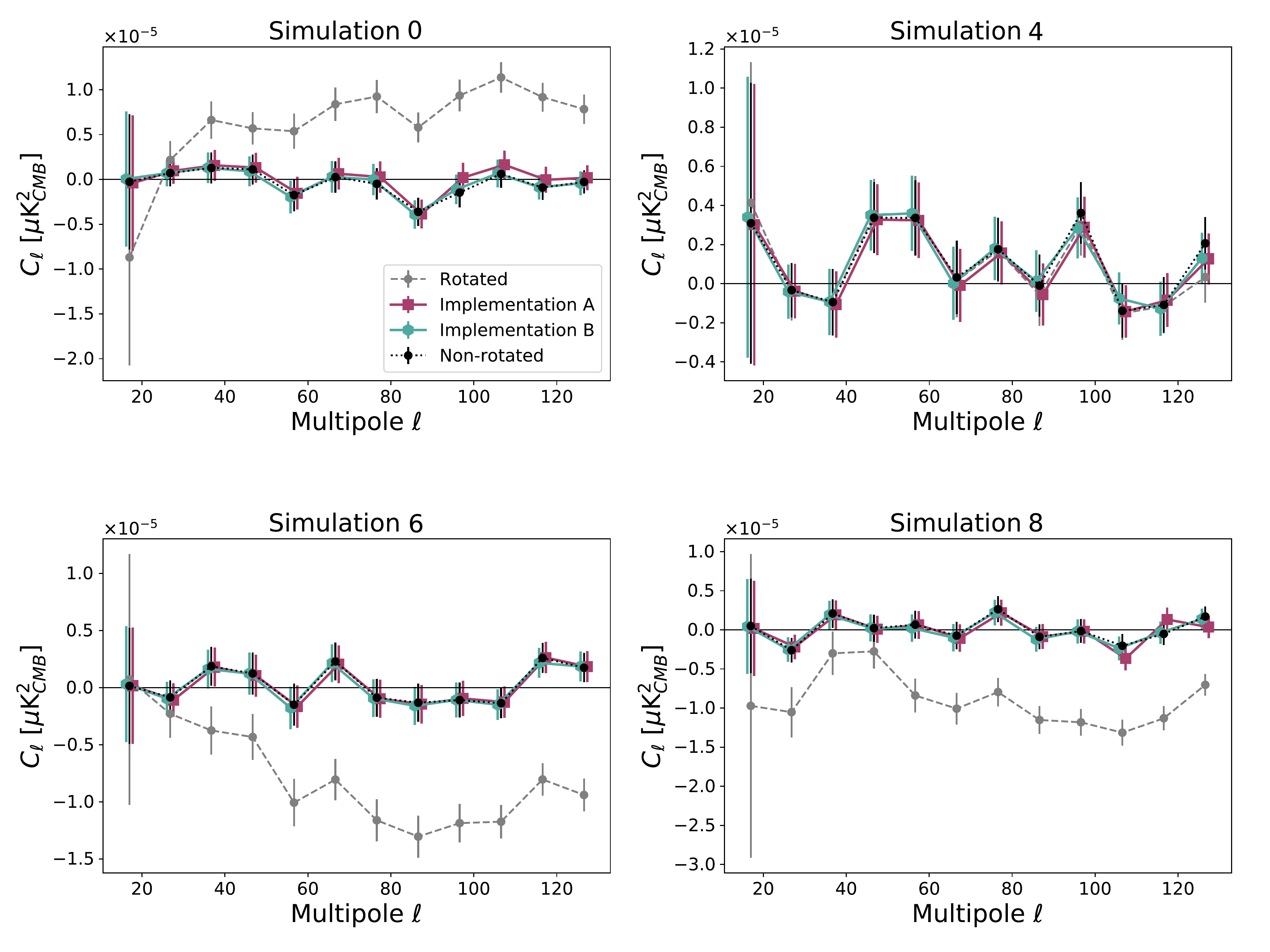}
    \caption {CMB $EB$ power spectra for all cases considered (non-rotated, rotated, and derotated maps with Implementations A and B),  after the application of the \bsecret\ algorithm to perform component separation. \rev{Note that, for visualization purposes, the green and purple points have been slightly shifted, to the left and right respectively, of the corresponding multipoles.}}
    \label{fig:cs_parametric_EB_power_spectra_residuals}
\end{figure}

Given the clean CMB map obtained using \bsecret, we calculate angular power spectra using a pseudo-$C_{\ell}$ estimator \citep{wandelt2001cosmic,2002ApJ...567....2H}, as implemented in the python \texttt{NaMaster} package \citep{alonso2019unified}. The power spectra are computed on 60\% of the sky defined by the publicly available \textit{Planck} mask\footnote{We use \texttt{HFI\_Mask\_GalPlane-apo0\_2048\_R2.00.fits} available in  \url{https://pla.esac.esa.int/##maps}}, in 10 multipole bins from $\ell=12$ to $\ell=132$. We have applied a non-uniform weighting scheme to account for the cosmic variance. Let $b$ be a bin, then the weight applied to $\ell_q \in b$ is
\begin{equation}
    w_{\ell_q} =  \dfrac{\ell_q+\frac{1}{2}}{\sum\limits_{\ell\in b} \left(\ell+\frac{1}{2}\right)} \, .
    \label{eq:weight_bins}
\end{equation}
Thus, the binned power spectrum is $C_b=\sum_{\ell\in b}w_\ell C_\ell$, and the binned covariance between the $i$-th and $j$-th bins is
\begin{equation}
    C_{i,j} = \sum\limits_{\ell_i\in b_i}\sum\limits_{\ell_j\in b_j} w_{\ell_i}w_{\ell_j}C_{\ell_i,\ell_j} \, ,
    \label{eq:binned_cov_mat}
\end{equation}
where $C_{\ell_i,\ell_j}$ is the covariance between the $i$-th and $j$-th multipoles.

\begin{figure}[t]
   \centering
\includegraphics[width=\textwidth]{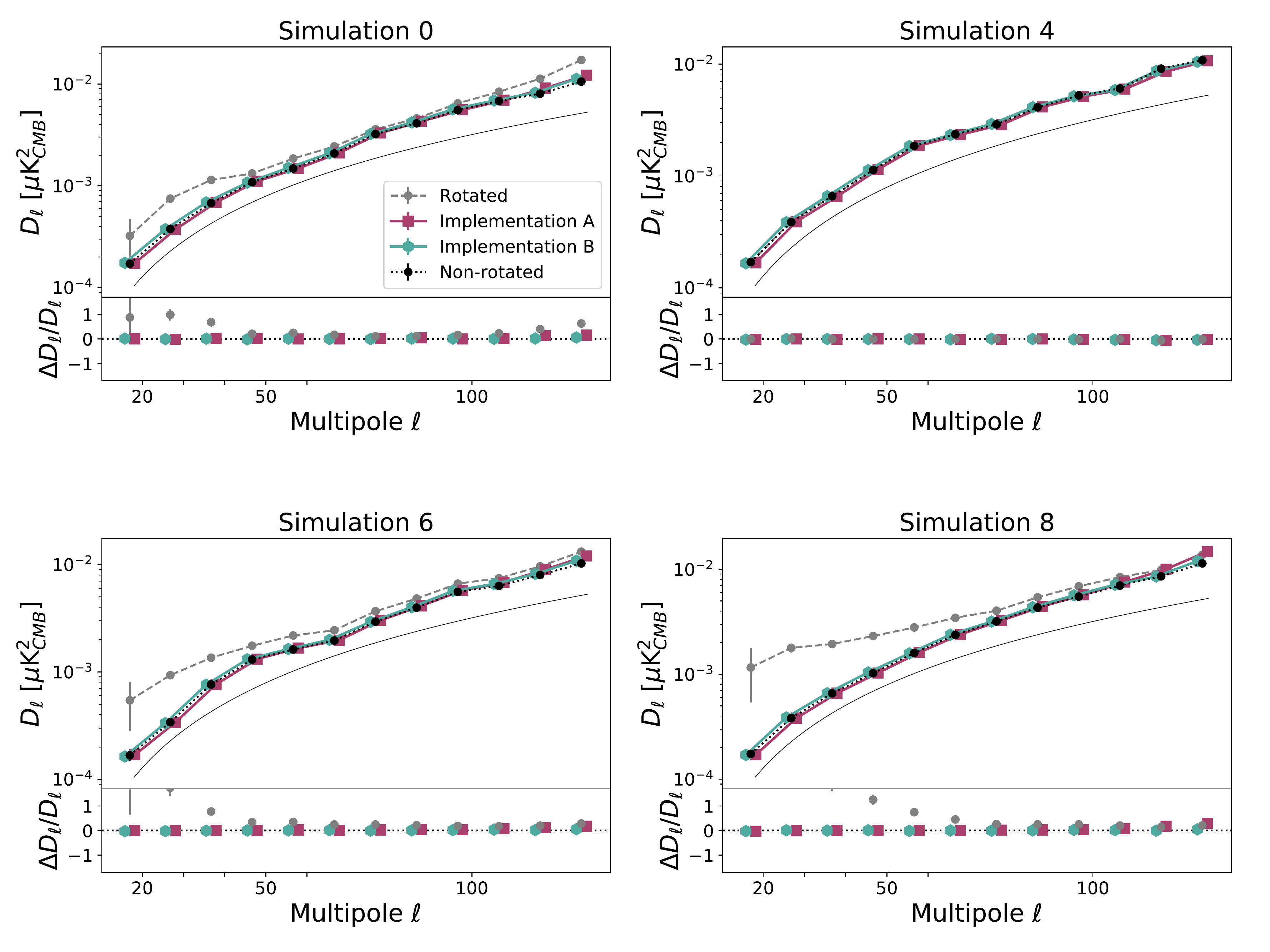}
    \caption{CMB $BB$ power spectra for the non-rotated, rotated, and derotated maps derived by Implementations A and B after application of the \bsecret\ component separation algorithm. The solid black line in the upper section of each panel shows the theoretical CMB lensing $BB$ power spectrum, $L_\ell$. The difference between $L_\ell$ and the data for non-rotated and derotated maps is due to the noise power spectrum, while that for rotated maps is also due to the component separation residual. The lower sections show the relative difference between the recovered $BB$ power spectra in the rotated and derotated cases with respect to the non-rotated case. \rev{Note that, for visualization purposes, the green and purple points have been slightly shifted, to the left and right respectively, of the corresponding multipoles. }}
    \label{fig:cs_parametric_BB_power_spectra_residuals}
\end{figure}

The results of the component separation procedure are reported in 
figures~\ref{fig:cs_parametric_EB_power_spectra_residuals} and \ref{fig:cs_parametric_BB_power_spectra_residuals}, which show the $EB$ and $BB$ power spectra of the recovered CMB signal, respectively. 
As expected, figure~\ref{fig:cs_parametric_EB_power_spectra_residuals} shows a non-null $C_{\ell}^{EB}$ contribution in the rotated maps, in agreement with eq.~(\ref{eq:cleanEB}).
On the other hand, the $C_{\ell}^{EB}$ of the non-rotated and derotated maps are compatible with zero.

Figure~\ref{fig:cs_parametric_BB_power_spectra_residuals} shows the excess in the $BB$ power spectra in the rotated maps compared to those in the non-rotated ones, that do not include any primordial $B$-mode signal. As the rotated maps carry frequency-dependent polarization angle offsets, which are not taken into account in the parametric model, the component separation fails and yields large residuals. We find that the amount of systematic residuals left in the rotated maps are realization dependent. For example, the systematic residuals in the 4th realization are negligible, while those in the 8th realization are significant, especially at low multipoles. 

On the other hand, the $BB$ power spectra of the non-rotated and derotated maps are similar, showing that the in-flight calibration methods proposed here are able to remove the systematic errors induced by the miscalibration of the polarization angles. Note that the correction of the polarization angle systematics is achieved in all realizations, regardless of whether the systematics are large (8th realization) or small (4th). Thus, the main goal of the present study is achieved.

\begin{table}
    \centering
    \begin{tabular}{c|ccc|ccc|ccc}
    \hline\hline
    \noalign{\smallskip}
    \multicolumn{1}{c}{} &
     \multicolumn{3}{c}{} & 
     \multicolumn{3}{c}{Correction with} & \multicolumn{3}{c}{Correction with} \\
    \multicolumn{1}{c}{} &
     \multicolumn{3}{c}{Rotated} & \multicolumn{3}{c}{Implementation A} & \multicolumn{3}{c}{Implementation B} \\
     \noalign{\smallskip}
    \hline
    \noalign{\smallskip}\noalign{\smallskip}
    nsim & $r \times 10^4$ & $ \dfrac{r-r^\text{non}}{\sigma_r^\text{non}}$ & $\dfrac{\sigma_r}{\sigma_r^\text{non}}$ & $r \times 10^4$ & $\dfrac{r-r^\text{non}}{\sigma_r^\text{non}}$ & $\dfrac{\sigma_r}{\sigma_r^\text{non}}$ & $r \times 10^4$ & $\dfrac{r-r^\text{non}}{\sigma_r^\text{non}}$ & $\dfrac{\sigma_r}{\sigma_r^\text{non}}$ \\
    \noalign{\smallskip}
    \hline 
    \noalign{\smallskip}
  0000 &  10.14 &  0.80 & 1.46 & $-$18.53 & $-$2.02 & 1.03 &  $-5.62$ & $-0.75$ & 1.01 \\
 0001 &  81.91 &  8.37 & 1.39 & $-$29.14 & $-$2.16 & 1.02 & $-$14.26 & $-$0.75 & 0.98 \\
 0002 &  42.78 &  4.67 & 1.24 & $-$26.46 & $-$2.52 & 1.02 &  16.30 &  1.92 & 0.97 \\
 0003 & 100.18 & 11.35 & 1.42 &   8.37 &  1.28 & 0.95 &  13.78 &  1.88 & 0.94 \\
 0004 &  21.92 &  0.46 & 1.00 &  22.31 &  0.50 & 0.99 &  23.07 &  0.57 & 0.97 \\
 0005 & 139.48 & 14.99 & 1.23 &  21.84 &  2.56 & 0.98 &  22.76 &  2.66 & 0.94 \\
 0006 &  99.26 &  8.81 & 1.47 & $-$20.48 & $-$2.35 & 1.01 &  $-$5.71 & $-$0.97 & 0.99 \\
 0007 &  51.22 &  3.08 & 1.09 &   4.20 & $-$1.59 & 1.03 &  25.59 &  0.53 & 1.00 \\
 0008 & 246.19 & 24.06 & 1.84 & $-$38.68 & $-$3.85 & 1.06 &  $-$4.51 & $-$0.50 & 1.02 \\
 0009 &  26.35 &  1.99 & 1.31 & $-$16.42 & $-$2.28 & 1.09 & $-$21.71 & $-$2.81 & 1.06 \\
  \noalign{\smallskip}
    \hline\hline
    \end{tabular}
    \caption{Comparison of $r$ estimation after the application of the \bsecret\ algorithm to the rotated and derotated maps with respect to $r$ from the non-rotated maps. The quantity ${(r-r^{\rm non})}/{\sigma_r^{\rm non}}$ shows the  ``number of $\sigma$s'' of the difference with respect to the non-rotated case while ${\sigma_r}/{\sigma_r^{\rm non}}$ is the ratio of the uncertainty calculated for a given case with respect to the non-rotated case. We recall that the input non-rotated simulations do not include any CMB primordial $B$-mode signal.}
    \label{tab:comparison_r_nonrotated}
\end{table}

\begin{table}
    \centering
    \begin{tabular}{c|ccc|ccc|ccc}
    
    \hline\hline
    \noalign{\smallskip}
    \multicolumn{1}{c}{} &
     \multicolumn{3}{c}{} & 
     \multicolumn{3}{c}{Correction with} & \multicolumn{3}{c}{Correction with} \\
     \multicolumn{1}{c}{} &
     \multicolumn{3}{c}{Rotated} & \multicolumn{3}{c}{Implementation A} & \multicolumn{3}{c}{Implementation B} \\
    \noalign{\smallskip}
    \hline
    \noalign{\smallskip}\noalign{\smallskip}
    nsim & $r \times 10^4$ & $ \dfrac{r-r^\text{non}}{\sigma_r^\text{non}}$ & $\dfrac{\sigma_r}{\sigma_r^\text{non}}$ & $r \times 10^4$ & $\dfrac{r-r^\text{non}}{\sigma_r^\text{non}}$ & $\dfrac{\sigma_r}{\sigma_r^\text{non}}$ & $r \times 10^4$ & $\dfrac{r-r^\text{non}}{\sigma_r^\text{non}}$ & $\dfrac{\sigma_r}{\sigma_r^\text{non}}$ \\
    \noalign{\smallskip}
    \hline
    \noalign{\smallskip}
 $0$ & $-2.07$ & $-0.77$ & $1.04$ &  $-6.23$ & $-1.22$ &  $1.02$ &  $-7.17$ & $-1.32$ &  $1.02$ \\
 $1$ & $-1.77$ & $-1.07$ & $1.07$ &  $-4.42$ & $-1.35$ &  $1.00$ &  $-4.37$ & $-1.34$ &  $1.00$ \\
 $2$ & $19.28$ &  $2.28$ & $1.00$ &   $1.40$ &  $0.40$ &  $0.98$ &   $0.74$ &  $0.33$ &  $0.98$ \\
 $3$ & $16.98$ &  $1.68$ & $1.11$ &  $-0.48$ & $-0.18$ &  $1.01$ &  $-1.96$ & $-0.34$ &  1.01 \\
 4 & 18.79 &  2.99 & 1.02 &  13.05 &  2.38 &  1.01 &  10.84 &  2.14 &  1.01 \\
 5 & $-2.55$ & $-1.06$ & 1.28 & $-10.80$ & $-1.94$ &  1.00 & $-10.54$ & $-1.92$ &  1.00 \\
 6 &  3.67 &  0.38 & 1.13 &  $-1.49$ & $-0.17$ &  1.01 &  $-1.87$ & $-0.21$ &  1.01 \\
 7 & 32.74 &  3.37 & 1.01 &  16.83 &  1.67 &  1.00 &  16.85 &  1.67 &  1.00 \\
 8 & 18.68 &  2.17 & 1.13 &   4.05 &  0.61 &  1.00 &   2.76 &  0.47 &  1.00 \\
 9 & 33.31 &  3.52 & 1.11 &  21.53 &  2.27 &  0.99 &  20.50 &  2.16 &  0.99 \\
 \noalign{\smallskip}
    \hline
    \hline
    \end{tabular}
    \caption{ Same as table~\ref{tab:comparison_r_nonrotated} but for the \NILC\ algorithm.}
    \label{tab:comparison_nilc_r_nonrotated}
\end{table}  

In order to assess the biases introduced in the $r$ parameter by the residual systematic effect, we compare the recovered $r$ values from each set of maps to those from the non-rotated maps. We fit the cleaned CMB $BB$ power spectrum for each case ($C_\ell^{\rm out}$) to a linear combination of the theoretical primordial $B$-mode power spectrum ($B_\ell^{\rm gw}$), the lensing $B$-mode power spectrum ($L_\ell$) and the power spectrum of the component separation residuals plus noise of non-rotated maps $R^{\rm non}_\ell$ (parametrized by the $a_R$ coefficient):
\begin{equation}
    -2\ln \mathcal{L}(r,a_R) = \mleft(C_{b}^{\rm out} - r B^{\rm gw}_{b} - L_{b} - a_R  R^{\rm non}_{b} \mright)^{\sf T}\mymatrix{C}^{-1}_{b,b'}\mleft(C_{b'}^{\rm out} - r B^{\rm gw}_{b'}- L_{b'} - a_R  R^{\rm non}_{b'} \mright) \, ,
    \label{eq:loglikelihood_r}
\end{equation}
where $b$ denotes a bin and $\mymatrix{C}_{b,b'}$ is the covariance between the power spectra at the bins $b$ and $b'$. The covariance matrix is calculated using the \texttt{gaussian\_covariance} subroutine in the \texttt{Namaster} package. This subroutine calculates the covariance matrix of the pseudo-$C_{\ell}$ power-spectra estimated as $\mymatrix{C} = \mymatrix{K}^{-1}\widetilde{\mymatrix{C}} (\mymatrix{K}^{-1})^T$ where $\mymatrix{\widetilde{C}}$ is the covariance of the pseudo power-spectra and, $\mymatrix{K}$ is the mode coupling matrix \cite{efstathiou2004myths,couchot2017cosmology}. The results are shown in table~\ref{tab:comparison_r_nonrotated}. We find that the recovered $r$  of the derotated maps with Implementations A and B are consistent with those of the non-rotated maps. We also find that the uncertainty on the recovered $r$ is approximately the same. On the other hand, there is a significant bias in $r$ of the rotated maps with higher uncertainties in those simulations where the impact of the systematic effect is higher. This mismatch between the rotated and non-rotated results arises because the non-rotated residuals, $R^{\rm non}_\ell$, account only for foregrounds and noise residuals; thus, the excess in the $BB$ power spectrum due to the angle miscalibration leads to a bias in $r$. We conclude that correcting the polarization angle miscalibration {\it prior to} the parametric component separation is crucial for an unbiased inference of $r$.

\subsection{\NILC}
\label{sec:nilc}

 Next, we use the \NILC\ algorithm to obtain clean CMB maps. As was done previously in the \bsecret\ case, we perform component separation on the non-rotated, rotated, and derotated maps for the ten sets of simulations. Before applying the algorithm, the simulated sky maps are first convolved and deconvolved in harmonic space to a common angular resolution. Here, we adopt the smallest beam, i.e., $17.9'$, as the common resolution.

The \NILC\ algorithm, as currently implemented, is applicable to scalar fields on the sphere; thus, we construct sky maps of the $E$ and $B$ modes from the input maps of the $Q$ and $U$ Stokes parameters on the full sky. The \NILC\ weights used to combine the multi-frequency input data to determine the CMB signal are computed separately for the $E$- and $B$-mode sky maps. The derived full mission weights are also applied to the ``half-split maps'', two splits of a map sharing the same sky signal but with uncorrelated noise enhanced by a factor of $\sqrt{2}$ with respect to the sensitivity levels reported in 
table~\ref{tab:LB_instrument}. These are subsequently used for both the power spectrum and noise estimation. 

The needlet weights are mostly determined by the Galactic contamination, which dominates on large angular scales, and by the noise, which dominates on small angular scales. The reconstructed CMB $E$- and $B$-mode maps cannot be free of contamination by residual foregrounds and noise. Therefore, for further analysis, a set of conservative masks are derived from the variance of the residual foreground maps as follows. First, the variance of the residual foreground maps are smoothed with a $9^{\circ}$ Gaussian beam. We then set
thresholds appropriate for the desired sky fraction. We pick the mask with a $60\%$ sky fraction for the $B$-mode map. 

Given the CMB sky obtained using NILC, we calculate angular power spectra using a pseudo-$C_{\ell}$ estimator \citep{2002ApJ...567....2H,2004MNRAS.350..914C,2005ApJ...622..759G,2005MNRAS.358..833T}. Although the NILC weights are computed from full mission sky maps, the impact of the instrumental noise residuals on the angular power spectra is avoided by evaluating cross-power spectra of the NILC half-split maps. 
Each data point of the angular power spectra is then obtained from the average of all possible cross half-split angular power spectra. To compute the covariance of our measurements, we follow the method described in Ref.  \cite{2005MNRAS.358..833T}.

\begin{figure}[t]
   \centering
\includegraphics[width=\textwidth]{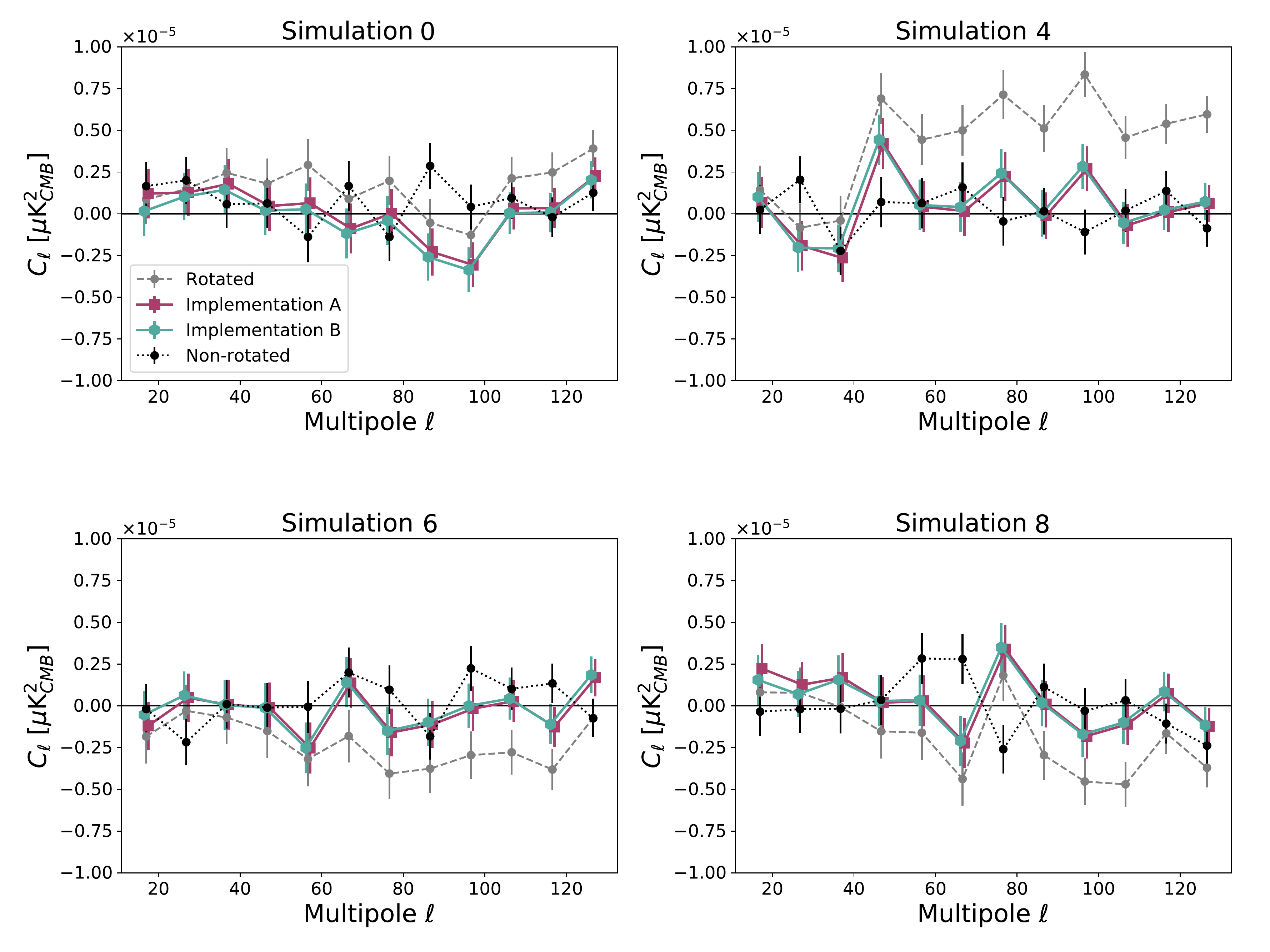}
    
    \caption{Same as figure~\ref{fig:cs_parametric_EB_power_spectra_residuals} but for the \NILC\ algorithm.}
    \label{fig:cs_nilc_EB_power_spectra}
\end{figure}

Figure~\ref{fig:cs_nilc_EB_power_spectra} shows the $EB$ power spectra of the clean CMB maps of the non-rotated, rotated, and derotated maps with Implementations A and B. 
We find that
the \NILC\ $EB$ power spectra of rotated maps are not consistent with a null result, in agreement with eq.~(\ref{eq:cleanEB}).
However they become compatible with zero in the derotated case (i.e., after correcting the polarization angle offsets), similar to the results of the parametric component separation. Note that, since we apply the correction to noisy maps, the noise realizations on the rotated and the derotated multi-frequency maps is different, and therefore the $EB$ spectra of the cleaned maps after the application of the \NILC\ algorithm are not expected to fully correlate, but only to be compatible with zero. 

\begin{figure}[t]
   \centering
\includegraphics[width=\textwidth]{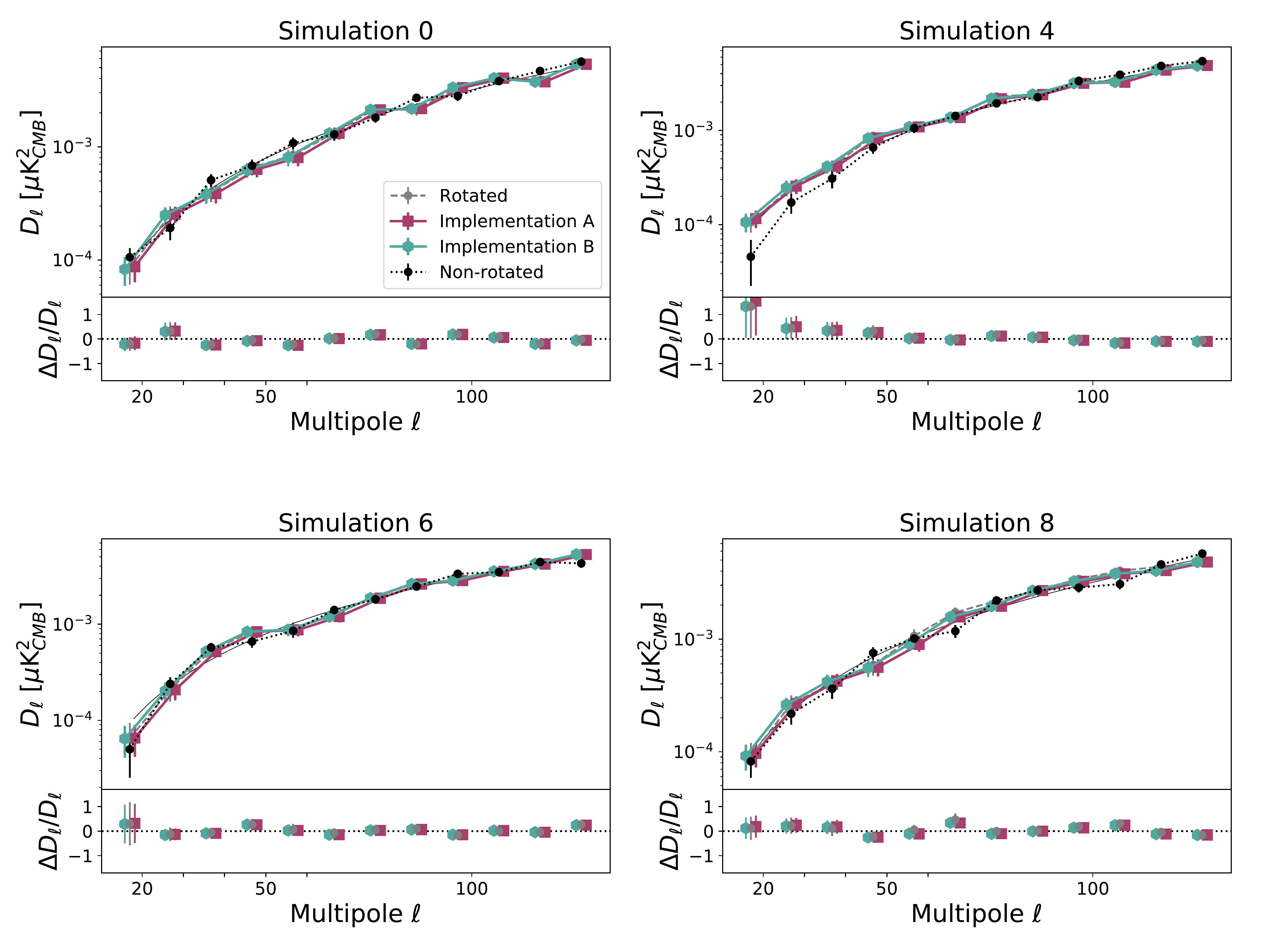}
    \caption{Same as figure~\ref{fig:cs_parametric_BB_power_spectra_residuals} but for the \NILC\ algorithm. Here, the noise power spectra are not seen because the $BB$ power spectra are estimated from half-split maps.}
    \label{fig:cs_nilc_BB_power_spectra}
\end{figure}

Interestingly, the $BB$ power spectra shown in figure~\ref{fig:cs_nilc_BB_power_spectra} are compatible with those of the non-rotated maps, as well as with the theoretical lensing $BB$ power spectrum $L_\ell$, not only 
for the derotated maps but also for the rotated maps.
This shows that, by being applied directly to the $B$-mode maps and by minimizing the variance of the CMB signal, the \NILC\ algorithm is also able to minimize the impact of the injected systematic effect. This property is fundamentally different from the outcome of the parametric component separation process.\par

Table~\ref{tab:comparison_nilc_r_nonrotated} reports the comparison of the estimation of $r$ for the different cases considered. The methodology to estimate $r$ and its error from the measured angular power spectra and covariance is the same as that for the paramateric component separation described in section~\ref{sec:pf}.  The multipole range and binning are also the same. As we estimate the power spectrum from half-split maps, we do not need to marginalize over the noise power spectrum; thus, we set $a_R=0$ in eq.~(\ref{eq:loglikelihood_r}).
A qualitatively different result compared to that from \bsecret\ is that the impact of the in-flight correction of the angle miscalibration is less evident, reflecting the fact that the CMB $BB$ power spectra do not show any excess due to the presence of the systematic effect. Nonetheless, the uncertainties on $r$ are reduced when the angle offset in the multi-frequency maps is corrected prior to the \NILC\ component separation.

\subsection{Impact on CMB maps}
\label{sec:impact_cmb}

We can further assess the impact of the polarisation angle offset and its residual before and after self-calibration by estimating the miscalibration angle $\alpha$ from the clean CMB maps. We estimate $\alpha$ by minimizing ${\cal D}_{\ell}^{EB}(\alpha)$ defined as
\begin{eqnarray}
{\cal D}_{\ell}^{EB}(\alpha) = C_{\ell}^{EB} \cos(4 \alpha) - \frac{1}{2} \left(C_{\ell}^{EE}-C_{\ell}^{BB}\right) \sin(4\alpha) \, ,
\label{DEB}
\end{eqnarray}
where $C_{\ell}^{EB}$, $C_{\ell}^{EE}$ and $C_{\ell}^{BB}$ are the power spectra of the CMB solutions, through a standard $\chi^2$-approach where the first $200$ multipoles are considered. The sky fraction used to evaluate the CMB power spectra is $f_{\rm sky} = 0.6$ for both component separation methods. We have also checked that the estimated angles are stable when we consider a smaller portion of the sky with $f_{\rm sky} )= 0.4$. Further details about the properties of ${\cal D}_{\ell}^{EB}(\alpha)$ can be found in Refs. \cite{Gruppuso:2016nhj,Aghanim:2016fhp}.

\begin{figure}[t]
\centering
\includegraphics[width=\textwidth]{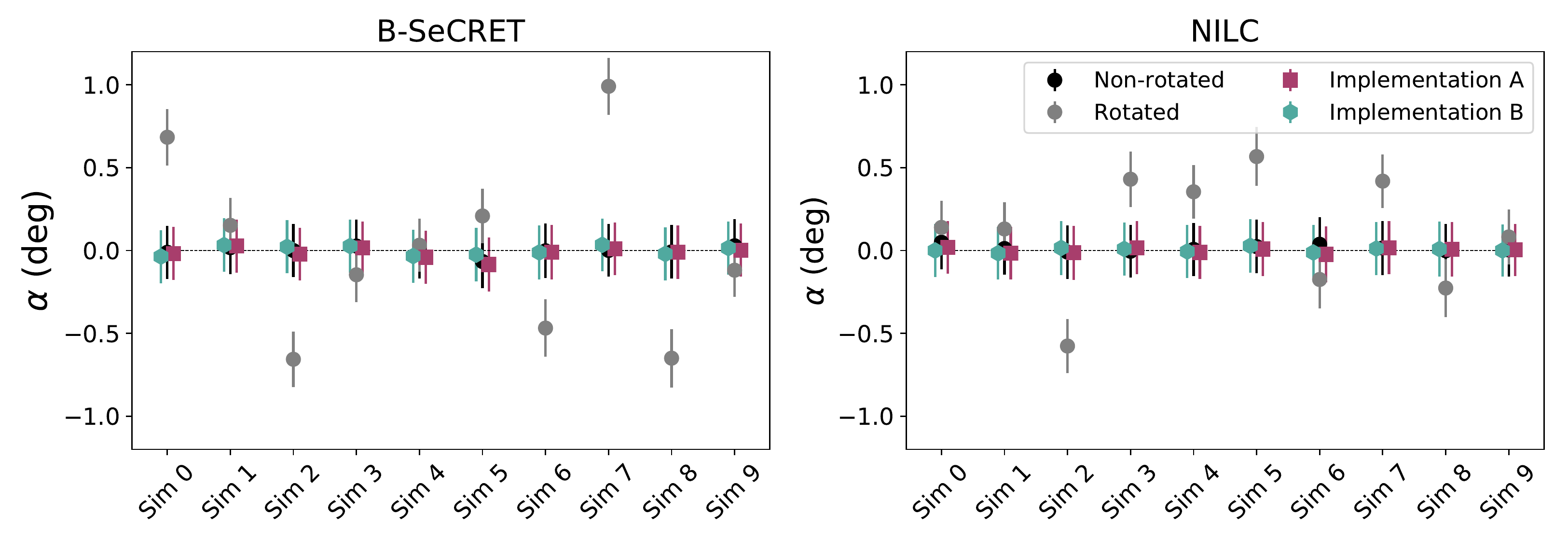}
\caption{Polarization angle estimated through minimization of $D_{\ell}^{EB}(\alpha)$ given in eq.~(\ref{DEB}), for each of the 10 realizations considered. This is performed on the CMB solutions found by \bsecret\ (left panel) and \NILC\ (right panel) algorithms. The black and gray points are for the non-rotated and rotated maps, respectively, whereas the purple and cyan points are for the derotated maps with Implementations A and B, respectively. The error bars are evaluated as the $99\%$ C.L. of $\exp{(-\chi^2/2)}$.}
\label{fig:angleresidual} 
\end{figure}

In figure~\ref{fig:angleresidual}, we show the angles $\alpha$  estimated from each of the ten realizations considered: the estimates from the non-rotated maps are shown in black, those from the rotated maps in grey, and those from the derotated maps are shown in purple for Implementation A and in cyan for B. The left and right panels show $\alpha$ of the CMB solutions found by \bsecret\ and \NILC\ algorithms, respectively.  
We find that $\alpha$ estimated from the \bsecret\ and \NILC\ rotated maps are different for most of the realizations. This is expected because the effective angles appearing in the $EB$ power spectrum of the clean CMB map ($\alpha_{{\rm eff},\ell}^E$ and $\alpha_{{\rm eff},\ell}^B$ in eq.~\ref{eq:cleanEB}) depend on the weights obtained for specific component separation algorithms.
The right panel shows that, although the angle offsets of the magnitudes assumed in this work do not lead to a large bias in $r$ from the \NILC\ CMB map, they still impact the $EB$ CMB cross-correlation, showing up with values of $\alpha$ significantly different from null, in agreement with eq.~(\ref{eq:cleanEB}).
On the other hand, when derotated with either Implementation A or B, we find excellent compatibility with zero for all realizations. \par 

\section{Discussion and conclusions}
\label{sec:discussion}
In this paper, we have presented the results of a blind analysis to study the impact of possible polarization angle offsets (miscalibration of linear polarization angles) on the measurements of the CMB polarized signal, with particular emphasis on the detection of the faint signal of the primordial $B$-mode polarization. We have used a set of simulated sky maps, where a rotation of the polarization angle was manually injected, generated by three different instrumental sources. As a test case we have considered the instrumental specification of the \LiteBIRD\ mission, with 22 (partially overlapping) frequency channels, each with a different polarization offset. As is well known, this systematic effect can cause a spurious $B$-mode signal, arising from the mixing of polarization states, with an amplitude potentially higher than the cosmological target range.\par

We have applied, in an independent manner, two different implementations of the self-calibration technique for correcting the polarization angle offset in each frequency channel, estimated by zeroing the $EB$ correlation. All of the frequency channels were analyzed jointly. The offset angles were recovered with an accuracy at the level of few arc-minutes. By propagating the residual angle miscalibration error to the component separation step, here represented by parametric fitting and ILC algorithms, we have estimated the impact on the measurement of the tensor-to-scalar ratio parameter, $r$. Results are reported in tables~\ref{tab:comparison_r_nonrotated} and~\ref{tab:comparison_nilc_r_nonrotated}. 

Table~\ref{tab:comparison_r_nonrotated} summarizes the impact of the angle miscalibration on $r$ using the parametric method \bsecret, showing that uncorrected offsets lead to a large bias in $r$. This is expected because the sky signal is rotated differently at each frequency channel and the parametric model of the sky emission is no longer adequate for the rotated maps. In the case of derotated maps, this effect is reduced significantly, the sky model is valid and $r$ from clean CMB maps of the derotated maps are consistent with those of the non-rotated maps. On the  other hand, table~\ref{tab:comparison_nilc_r_nonrotated} shows that the impact of the angle miscalibration on the angular power spectrum of CMB $B$ modes obtained using \NILC\ is small and is not prominently reflected in any bias on $r$. This is a consequence of the fact that the method makes no  assumptions about the foreground emission, but finds a solution by minimizing the variance of the clean CMB map. It is only mildly affected by incorrect modelling of the sky, and is capable of dealing with the extra complication introduced by this systematic. However, we still find an increase in the uncertainty on $r$ compared to the derotated cases. 

The additional analysis of the recovered CMB maps, presented in section~\ref{sec:impact_cmb} shows that the presence of residual polarization angle offsets is detectable as a non-zero signal in the clean CMB $EB$ power spectra, even in the \NILC\ case where the impact on the cosmological parameter $r$ is small. The spectra are found to be compatible with zero when corrections are applied.

In conclusion, we have shown how the interplay between errors on the calibration of the instrumental polarization angles and component separation method can lead to a bias in $r$ when clean CMB maps are retrieved with parametric methods without correcting the angle miscalibration.
Component separation methods that do not make any assumptions about the foreground emission, such as \NILC, are less affected by miscalibration. In both cases, the clean CMB maps of rotated maps show non-null $EB$ power spectra.
The $EB$ self-calibration method to correct the polarization angle offsets can efficiently restore the correct instrument orientation and mitigate their impacts,  if applied prior to the component separation (especially for the parametric method, but also for non-parametric ones, leading to smaller uncertainties).

\acknowledgments
This work is supported in Japan by ISAS/JAXA for Pre-Phase A2 studies, by the acceleration program of JAXA research and development directorate, by the World Premier International Research Center Initiative (WPI) of MEXT, by the JSPS Core-to-Core Program of A. Advanced Research Networks, and by JSPS KAKENHI Grant Numbers JP15H05891, JP17H01115, and JP17H01125. 
The Italian LiteBIRD phase A contribution is supported by the Italian Space Agency (ASI Grants No. 2020-9-HH.0 and 2016-24-H.1-2018), the National Institute for Nuclear Physics (INFN) and the National Institute for Astrophysics (INAF). 
The French LiteBIRD phase A contribution is supported by the Centre National d’Etudes Spatiale (CNES), by the Centre National de la Recherche Scientifique (CNRS), and by the Commissariat à l’Energie Atomique (CEA). 
The Canadian contribution is supported by the Canadian Space Agency. The US contribution is supported by NASA grant no. 80NSSC18K0132. 
Norwegian participation in LiteBIRD is supported by the Research Council of Norway (Grant No. 263011). 
The Spanish LiteBIRD phase A contribution is supported by the Spanish Agencia Estatal de Investigación (AEI), project refs. PID2019-110610RB-C21 and AYA2017-84185-P. 
Funds that support the Swedish contributions come from the Swedish National Space Agency (SNSA/Rymdstyrelsen) and the Swedish Research Council (Reg. no. 2019-03959). 
The German participation in LiteBIRD is supported in part by the Excellence Cluster ORIGINS, which is funded by the Deutsche Forschungsgemeinschaft (DFG, German Research Foundation) under Germany’s Excellence Strategy (Grant No. EXC-2094 - 390783311). 
This research used resources of the Central Computing System owned and operated by the Computing Research Center at KEK, as well as resources of the National Energy Research Scientific Computing Center, a DOE Office of Science User Facility supported by the Office of Science of the U.S. Department of Energy.
TM's work is supported by JSPS KAKENHI Grant Number JP18KK0083.
EdlH acknowledges partial financial support from the \textit{Concepci\'on Arenal Programme} of the Universidad de Cantabria. EdlH, EMG and PV acknowledge the Santander Supercomputaci\'on support group at the Universidad de Cantabria, a member of the Spanish Supercomputing Network, who provided access to the Altamira Supercomputer at the Instituto de F\'isica de Cantabria (IFCA-CSIC) for performing simulations and analyses. 
They also acknowledge funding from Unidad de Excelencia Mar{\'\i}a de Maeztu (MDM-2017-0765).
YM's work was supported in part by the Japan Society for the Promotion of Science (JSPS) KAKENHI, Grants No.~JP20K14497. 
NK, CB and AG acknowledge financial support from the INFN InDark project and from the COSMOS network (www.cosmosnet.it) through the ASI (Italian Space Agency) Grants 2016-24-H.0 and 2016-24-H.1-2018.
EK's work was supported in part by 
the JSPS KAKENHI Grant Number JP20H05850 and JP20H05859.

\appendix
\section{Methods to estimate the polarization angle offset}
In section~\ref{Sec:methods}, we provided a high-level summary of two implementations of the self-calibration technique for estimating the polarization angle offsets from the multi-frequency simulated maps, which we called ``Implementation A'' and ``Implementation B''. Here, we describe the details. 
\label{App:pol_methods}
\subsection{Implementation A}
\label{sec:A}
This is a computationally fast implementation of the methodology to estimate the parameter $\alpha$ from the observed power spectra. 
We make two assumptions that enable us to obtain analytic formulae for both the rotation angles and their uncertainties in terms of the $EE$, $BB$ and $EB$ power spectra. The formalism of this implementation is explained in \cite{de2021determination}. We build our likelihood function considering only the auto-frequency power spectra, which are computed on the full sky with the \texttt{anafast} function within the \texttt{healpy} library.  Despite this limitation in the information used, the accuracy of the recovered polarized angles is sufficiently competitive. 

The main assumptions made for simplifying the likelihood are the following:
\begin{itemize}
    \item Small angle approximation: $\tan(\alpha) \to \alpha$.
    \item We do not vary $\alpha$ in the covariance matrix of the power spectra when estimating $\alpha$. To correct the mismatch induced by this approximation, we perform an iterative approach that updates $\alpha$ in the covariance matrix with the one estimated in the previous step.
\end{itemize}
 We ignore correlations between different multipoles in the likelihood, since we work with the full-sky data and thus the correlation is negligible.
With these approximations, we obtain a linear system from which the analytical equations to calculate the rotation angles as well as the Fisher error bars are derived. The  analyticity of the problem yields a fast computational implementation. 

After applying the aforementioned approximations, the likelihood is given by
\begin{equation}
    -2 \ln \mathcal{L} = \sum\limits_{i=1}^{N_{\rm ch}}  \sum\limits_{j=1}^{N_{\rm ch}}  \sum\limits_{\ell=\ell_{\rm min}}^{\ell_{\rm max}} \left(C_{\ell}^{E_iB_i,{\rm o}} - 4\alpha_i\xi_{\ell}^{i}\right)\left(\mymatrix{M} ^{-1}\right)_{\ell,ij}
    \left(C_{\ell}^{E_jB_j,{\rm o}} - 4\alpha_j\xi_{\ell}^{j}\right)\, ,
    \label{eq:auto_likelihood}
\end{equation}
where $N_{\rm ch}$ is the number of frequency channels and  $\alpha_i$ is the $i$-th channel's polarization angle offset. We find that $\ell_{\rm min}=10$ and $\ell_{\rm max}=300$ are the optimal multipole range for a \LiteBIRD-like instrument.
Here, $\xi_{\ell}^{i}$\footnote{Notice that this is valid only if the noise bias is null. Otherwise, the noise bias could be taken into account.} is given by
\begin{equation}
    \xi_{\ell}^{i} = \frac12\left({C_{\ell}^{E_iE_i,{\rm o}} - C_{\ell}^{B_iB_i,{\rm o}}}\right) \, .
    \label{eq:Delta_likelihood}
\end{equation}
The elements of the covariance matrix $\mymatrix{M}$ are given by
\begin{align}
    M_{\ell,ij} = \dfrac{1}{2\ell +1} 
     & \left\{\ensemble{C}_{\ell}^{E_iE_j,{\rm o}}\ensemble{C}_{\ell}^{B_iB_j,{\rm o}} + \ensemble{C}_{\ell}^{E_iB_j,{\rm o}}\ensemble{C}_{\ell}^{B_iE_j,{\rm o}}\right.\nonumber \\  
     & - 4\alpha_j\left(\ensemble{C}_{\ell}^{E_iE_j,{\rm o}}\ensemble{C}_{\ell}^{B_iE_j,{\rm o}} - \ensemble{C}_{\ell}^{E_iB_j,{\rm o}}\ensemble{C}_{\ell}^{B_iB_j,{\rm o}}\right) \nonumber \\
    & - 4\alpha_i\left(\ensemble{C}_{\ell}^{E_iE_j,{\rm o}}\ensemble{C}_{\ell}^{E_iB_j,{\rm o}} - \ensemble{C}_{\ell}^{B_iE_j,{\rm o}}\ensemble{C}_{\ell}^{B_iB_j,{\rm o}}\right)\nonumber \\
    & \left . + 8\alpha_i\alpha_j\left[\left(\ensemble{C}_{\ell}^{E_iE_j,{\rm o}}\right)^2 + \left(\ensemble{C}_{\ell}^{B_iB_j,{\rm o}}\right)^2 -\left(\ensemble{C}_{\ell}^{E_iB_j,{\rm o}}\right)^2 - \left(\ensemble{C}_{\ell}^{B_iE_j,{\rm o}}\right)^2 \right]\right\}  \, ,
    \label{eq:auto_covariance_matrix}
\end{align}
where $\ensemble{C}_{\ell}^{XY}$ are the power spectra smoothed by convolving $C_{\ell}^{XY}$ with a 5-$\ell$ width box of unity area.

The rotation angles can then be obtained analytically by solving the following linear system,
\begin{equation}
    \sum_{j=1}^{N_{\rm ch}}\Omega_{ij}\alpha_j = \dfrac{1}{4}\eta_i \, ,
    \label{eq:auto_angles_linear_system}
\end{equation}
where 
\begin{align}
    \Omega_{ij} &= \sum\limits_{\ell=\ell_{\rm min}}^{\ell_{\rm max}} \xi_{\ell}^{i}\left(\mymatrix{M} ^{-1}\right)_{\ell,ij}\xi_{\ell}^{j} \, , 
    \label{eq:omega_element} \\
    \eta_{i} &= \sum\limits_{j=1}^{N_{\rm ch}}\sum\limits_{\ell=\ell_{\rm min}}^{\ell_{\rm max}} \xi_{\ell}^{i}\left(\mymatrix{M} ^{-1}\right)_{\ell,ij}C_{\ell}^{E_jB_j} \, .
    \label{eq:eta_element}
  \end{align} 
In this implementation, the uncertainties on the rotation angles are obtained from the Fisher matrix, whose elements are given by
\begin{equation}
    F_{ij} = \dfrac{1}{2}\dfrac{\partial^2 (-2\ln \mathcal{L})}{\partial \alpha_{i}\partial \alpha_{j}} = 16 \sum\limits_{\ell=\ell_{\rm min}}^{\ell_{\rm max}} \xi_{\ell}^{i}\left(\mymatrix{M} ^{-1}\right)_{\ell,ij}\xi_{\ell}^{j}\, .
    \label{eq:Fisher_auto_elements}
\end{equation}

\subsection{Implementation B}
\label{sec:B}

In this implementation, 
we extend eq.~\eqref{eq:EBobs} to include all the possible observed $EB$ cross power spectra from $N_\mathrm{ch}$ frequency channels to estimate $\alpha_i$, allowing us to retrieve the values of the parameters with more precision.
We include the approximate covariance between all the observed $EB$, $EE$, and $BB$ power spectra assuming Gaussian statistics.
The methodology used for this implementation and its validation are detailed in Ref. \cite{minami/komatsu:2020a}.
We briefly review the methodology below.

When we ignore the intrinsic $EB$ cross power spectra of the CMB and the Galactic foregrounds,
we can relate the observed power spectra of the $i$th and $j$th channels at each $\ell$ as~\cite{minami/komatsu:2020a},
\begin{align}\label{eq:CellObs}
\begin{pmatrix}
-\vec{R}^{\sf T} (\alpha_i, \alpha_j)\mathbf{R}^{-1}(\alpha_i, \alpha_j)  & 1
\end{pmatrix}
   \begin{pmatrix}
   C_\ell^{E_i E_j, \mathrm{o}}   \\ C_\ell^{B_i B_j, \mathrm{o}} \\   C_\ell^{E_i B_j, \mathrm{o}}
   \end{pmatrix}
   = 0,
\end{align}
where $\mathbf{R}$  and $\vec{R}$ are a rotation matrix and rotation vector of power spectra, respectively.
The explicit forms are 
\begin{align}
\mathbf{R} 
( \theta_i ,  
\theta_j ) &= \begin{pmatrix}
\cos(2\theta_i) \cos(2\theta_j) & \sin(2\theta_i) \sin(2\theta_j)
\\
\sin(2\theta_i) \sin(2\theta_j) & \cos(2\theta_i) \cos(2\theta_j)
\end{pmatrix}\,,\\
\vec{R} (\theta_i , \theta_j)&= \begin{pmatrix}
\cos(2\theta_i)\sin(2\theta_j) 
\\
-\sin(2\theta_i) \cos(2\theta_j)
\end{pmatrix}\,.
\end{align}

Using eq.~(\ref{eq:CellObs}), 
we construct a log-likelihood function as~\cite{minami/komatsu:2020a},
\begin{align}\label{eq:LikelihoodCross}
-2\ln\mathcal{L} = 
\sum_{\ell=\ellmin}^{\ellmax}
\left( \mathbf{A}\vec{C}_\ell^{\mathrm{o}} \right)^{\sf T} 
\mathbf{C}^{-1}
\left( \mathbf{A}\vec{C}_\ell^{\mathrm{o}}\right)\,,
\end{align}
where $\ellmin=2$, $\ellmax=1024$,
$\vec{C}_\ell^{\mathrm{o}}$ is an array of the observed power spectra,
$\begin{pmatrix}
C_\ell^{E_i E_j, \mathrm{o}}   & C_\ell^{B_i B_j, \mathrm{o}} &   C_\ell^{E_i B_j, \mathrm{o}}
\end{pmatrix}^{\sf T}$, with $i,j$ in $_{22}\mathrm{C}_{2} + 22 = 253$ combinations,
$\mathbf{A}$ is a block diagonal matrix of $\begin{pmatrix}
-\vec{R}^{\sf T} (\alpha_i, \alpha_j)\mathbf{R}^{-1}(\alpha_i, \alpha_j)  & 1
\end{pmatrix}$, and 
$\mathbf{C} = \mathbf{A}\Cov(\vec C_\ell^{\mathrm{o}},\vec C_\ell^{\mathrm{o}}{}^{\sf T})\mathbf{A}^{\sf T}$.
The explicit form of $\Cov(\vec C_\ell^{\mathrm{o}},\vec C_\ell^{\mathrm{o}}{}^{\sf T})$ is 
\begin{equation}\label{eq:Cov}
\begin{split}
&\Cov(\vec{C}_\ell^{\mathrm{o}, ij}, \vec C_\ell^{\mathrm{o},pq}{}^{\sf T}) \\
&=
 \begin{pmatrix}
 \Cov( C_\ell^{E_i E_j,\mathrm{o}},  C_\ell^{E_p E_q,\mathrm{o}})&\Cov( C_\ell^{E_i E_j,\mathrm{o}},  C_\ell^{B_p B_q,\mathrm{o}})&\Cov( C_\ell^{E_i E_j,\mathrm{o}}, C_\ell^{E_p B_q,\mathrm{o}})\\
 \Cov( C_\ell^{B_i B_j,\mathrm{o}}, C_\ell^{E_p E_q,\mathrm{o}})&\Cov( C_\ell^{B_i B_j,\mathrm{o}}, C_\ell^{B_p B_q,\mathrm{o}})&\Cov( C_\ell^{B_i B_j,\mathrm{o}}, C_\ell^{E_p B_q,\mathrm{o}})\\
 \Cov( C_\ell^{E_i B_j,\mathrm{o}}, C_\ell^{E_p E_q,\mathrm{o}})&\Cov( C_\ell^{E_i B_j,\mathrm{o}}, C_\ell^{B_p B_q,\mathrm{o}})&\Cov( C_\ell^{E_i B_j,\mathrm{o}}, C_\ell^{E_p B_q,\mathrm{o}})
 \end{pmatrix},
\end{split}
\end{equation}
where we use an approximate covariance for each element~\cite{minami/komatsu:2020a}:
\begin{align}
\begin{split}
\Cov(C_\ell^{X,Y,\mathrm{o}}, C_\ell^{Z,W,\mathrm{o}}) &
\approx \frac{1}{(2\ell+1) }(C_\ell^{X,Z,\mathrm{o}}C_\ell^{Y,W,\mathrm{o}} + C_\ell^{X,W,\mathrm{o}} C_\ell^{Y,Z,\mathrm{o}}).
\end{split}
\end{align}
Thus,
we estimate $\alpha_i$ only with the observed power spectra.
We do not include $\ln| \bf{C} |$ term in the log-likelihood function of eq.~\eqref{eq:LikelihoodCross}, following the method validated in Ref. \cite{minami/komatsu:2020a}. We confirm that including the $\ln| \bf{C} |$ term does not change the results for the instrument specification given in table~\ref{tab:LB_instrument}.

Here, to remove biases from statistical fluctuations,
we neglect the off-diagonal elements in eq.~(\ref{eq:Cov})
and adopt the binned power spectra and the corresponding covariance~\cite{minami/komatsu:2020a}:
\begin{align}
\begin{split}
C_b^{X,Y} &= \frac{1}{\Delta\ell}\sum_{\ell \in b} C_\ell^{X,Y}\,,\\
\Cov(C_b^{X,Y}, C_b^{Z,W}) &= \frac{1}{\Delta\ell^2}\sum_{\ell\in b}\Cov(C_\ell^{X,Y}, C_\ell^{Z,W})\,,
\end{split}
\end{align}
where we use $\Delta\ell = 20$.
Using these binned variables in the log-likelihood ~(\ref{eq:LikelihoodCross}) 
we estimate all $\alpha_i$ by sampling the posterior distribution with the MCMC algorithm \textsc{emcee}~\cite{foreman2013emcee} and evaluating the errors on the retrieved maximum likelihood $\alpha_i$ parameters at $1\sigma$.

\section{Component Separation Methodologies}
\label{app:component_separation_methodologies}
In this section, we describe the component separation methods \bsecret\ \cite{de2020detection} and \NILC\ \cite{2009A&A...493..835D, 2012MNRAS.419.1163B,2013MNRAS.435...18B} adopted to obtain clean CMB maps from the multi-frequency simulated sky maps.

\subsection{\bsecret}
\label{app:Bsecret}
The first component separation method is based on a modified version of the Bayesian parametric pixel-based maximum likelihood method described in \citep{de2020detection}. 
Given a set of sky maps observed at different frequencies, the multi-frequency signal at a given pixel $p$ is fitted to a parametric model given by 
\begin{equation}
    m^{\notsotiny{X}}\mleft(\nu;\myset{\theta}_{p}^{\notsotiny{X}}\mright) = c_{p}^{\notsotiny{X}} + \dfrac{a_{\text{s},p}^{\notsotiny{X}}}{u(\nu)}\mleft(\dfrac{\nu}{\nu_\text{s}}\mright)^{\beta_{\text{s}}+c_{\text{s}}\log\mleft(\nu/\nu_{\text{s}}\mright)} + \dfrac{a_{\text {d},p}^{\notsotiny{X}}}{u(\nu)}\mleft(\dfrac{\nu}{\nu_\text{d}}\mright)^{\beta_{\text{d}}-2}
    \dfrac{B\mleft(\nu,T_{\text {d}}\mright)}{B\mleft(\nu_{\text {d}},T_{\text{d}}\mright)}
    \,,
    \label{eq:cs_parametric_model}
\end{equation}
where $X$ denotes a Stokes parameter ($Q$ or $U$).
We  use an MCMC algorithm \textsc{emcee}~\cite{foreman2013emcee} to sample the posterior of $\theta_p^{\notsotiny{X}} = \{c_p^{\notsotiny{X}},a_{\text s,p}^{\notsotiny{X}},a_{\text d,p}^{\notsotiny{X}},\beta_{\text s},\beta_{\text d},c_{\text s},T_{\text d}\}$. Here, $u(\nu) = x^2e^x/(e^x-1)^2$ with $x =h\nu/(k_{\text  B}T_{\text CMB})$ is a unit conversion factor from thermodynamic to antenna temperature units, $\nu_{\text s} = \nu_{\text d} = 150$GHz are the synchrotron and dust pivot frequencies, and $B(\nu,T)$ is Planck's law. The second and third terms on the right hand side are the synchrotron and dust contributions, respectively. We use Gaussian priors for the foreground SED parameters: $\beta_{\text s}\sim \mathcal{N}(-3.1,0.3)$, $\beta_{d}\sim \mathcal{N}(1.56,0.1)$, $c_{\text s}\sim \mathcal{N}(0,0.1)$, and $T_{\text d}\sim \mathcal{N}(21,3)$.

The posterior probability density is given by the product of the data likelihood and the priors on the model parameters. The likelihood for each pixel $p$ is given by
\begin{equation}
    \mathcal{L}(\myset{\theta}_{p}|\myvector{d}_{p}) = 
    \dfrac{1}{\sqrt{(2\pi)^{2N_{\rm ch}}\det (\mymatrix{C})}}
    \exp\mleft[-\dfrac{1}{2}\mleft(\myvector{d}_{p}-m\mleft(\myvector{\nu};\myset{\theta}_{p}\mright)\mright)^{\sf T}\mymatrix{C}^{-1}\mleft(\myvector{d}_{p}-m\mleft(\myvector{\nu};\myset{\theta}_{p}\mright)\mright)\mright] \,,
    \label{eq:likelihood}
\end{equation}
where $N_{\rm ch}=22$ is the number of frequency channels, $\myvector{\nu}$ is a $N_{\rm ch}$ vector whose elements are \textit{LiteBIRD}'s frequency channels, $\myvector{d_p} = (\myvector{d_p}^{Q},\myvector{d_p}^{U})$ with $\myvector{d_p}^{Q}$ ($\myvector{d_p}^{U}$) being a $N_{\rm ch}$ vector with the $Q$ ($U$) multi-frequency signal in the pixel $p$, $\myset{\theta}_p$ is the set of model parameters in the pixel $p$,  $m(\myvector{\nu},\theta_p) =(m^{Q}(\myvector{\nu},\theta_p),m^{U}(\myvector{\nu},\theta_p))$ with $m^{Q}(\myvector{\nu},\theta^{Q}_p)$ ($m^{U}(\myvector{\nu},\theta^{U}_p)$) being a vector whose elements are the result of evaluating the model given the parameters $\theta^{Q}_p$ ($\theta^{U}_p$) in the pixel $p$, and $\mymatrix{C} = \text{diag}(\mathbf{C}^{\textrm{\notsotiny{Q}}},\mymatrix{C}^{\notsotiny{U}})$  with $\mymatrix{C}^{\notsotiny{X}}$ being a $N_{\rm ch}\times N_{\rm ch}$ noise covariance matrix for the $X$ Stokes parameter. We assume $\mymatrix{C}^{\notsotiny{Q}} = \mymatrix{C}^{\notsotiny{U}}$. 
The matrix $\mymatrix{C}$ depends in general on $p$, but we assume it to be independent of it, i.e., homogeneous noise across the sky. $\mymatrix{C}^{\notsotiny{X}}$ is assumed to be diagonal, i.e., no correlation among different frequencies.  The $Q$ and $U$ signals are fitted jointly since we assume that they share the same SED model parameters. Since the foregrounds have been simulated using the uniform SED parameters, we also assume uniform spectral parameters across the available sky.

Due to the large computational time required to fit the maps at $N_{\rm side}=512$, we downgrade them to $N_{\rm side}=64$ and convolve them to a common beam resolution. The map processing proceeds as follows: (i) the original maps at $N_{\rm side}=512$ are converted to a spherical harmonic representation; (ii) beam deconvolution is applied in the harmonic domain for the beam full-width-at-half-maximum (FWHM) of each frequency channel as reported in table~\ref{tab:LB_instrument}; (iii) the spherical harmonic coefficients of each channel are convolved with a Gaussian beam of FWHM$=132'$; and (iv) we transform the spherical harmonic coefficients to $N_{\rm side}=64$ maps. 

To estimate the effective noise per channel, we have generated 100 noise simulations per \LiteBIRD\ channel and downgraded them using the same process. The $i$-th diagonal element of $\mymatrix{C}^{\notsotiny{X}}$ is the effective variance calculated from the 100 pre-processed noise simulation maps.

We perform the parameter fitting in a two-step process, as in Ref. \citep{de2020detection}. 
First, the model parameters are split into two categories: the amplitudes and the SED parameters. Each set of the parameters is fitted in an iterative manner, i.e., in the first iteration the SED parameters are fixed to the initial values of $\beta_{\text s} =-3.0$, $\beta_{\text d}=1.54$, $c_{\text s} = 0$ and $T_d=20$~K, and the amplitudes are fitted. Then, the amplitudes are fixed to the values obtained from this fit and the SED parameters are calculated. In the next step, the SED parameters are fixed to the values determined in the first iteration, and the amplitudes are fitted. This process is repeated until it converges. We find that convergence is achieved typically by the second iteration.

\subsection{\NILC}
\label{sec:nilc_method}
The second component separation technique that we apply, \NILC\ \citep{2009A&A...493..835D, 2012MNRAS.419.1163B,2013MNRAS.435...18B}, is based on the ILC method \cite{1996MNRAS.281.1297T,Bennett:2003ca,2003PhRvD..68l3523T}. It is based on the construction of the linear mixture of frequency channel maps that minimizes the variance on a frame of spherical wavelets called needlets, allowing localised filtering in both pixel space and harmonic space. It is designed to recover the CMB as the component scaling as a blackbody in the linear mixture, assuming only that it is uncorrelated with foregrounds, with no other prior information. See Refs. \cite{2016A&A...594A...9P,Planck2018_IV} for the application to the \textit{Planck} data.

\NILC  estimates the CMB, ${\widehat S}$, as a weighted linear combination of multi-frequency sky maps such that (1) the variance of the estimate is minimum, with (2) unit response to the flat CMB frequency spectrum,
\begin{eqnarray} 
\widehat S = w^{\sf T} X = \frac{\displaystyle{a^{\sf T} \widehat R^{-1}}}{\displaystyle{a^{\sf T} \widehat R^{-1} a}} X =\frac{\displaystyle{a^{\sf T} \widehat R^{-1}}}{\displaystyle{a^{\sf T} \widehat R^{-1} a}} \left(a^{ } S + F + N\right)\,.
\label{equ:ilc} 
\end{eqnarray}
Here, $X$ is the vector of frequency channel maps, $a$ the constant frequency spectrum of the CMB signal $S$, $F$ the total foreground signal, $N$ the instrumental noise for the different frequency channels, and ${\widehat R}$ the covariance matrix across frequencies. The  condition (1) guarantees minimum contamination by foregrounds and instrumental noise, while condition (2) guarantees that the CMB signal is conserved without bias. The weights, $w$, result from a trade-off between minimising the foregrounds and the instrumental noise contributions in the reconstructed CMB map \citep{2006NewAR..50..854S,2009A&A...493..835D,2008PhRvD..78b3003S,2011ApJ...739L..56S,2010MNRAS.401.1602D}. They are computed in needlet space, i.e., for different regions of the sky or for different angular scales, respectively, which allows for variations of the data covariance matrix in either space. This technique has already been applied broadly in CMB data analysis \citep{2009A&A...493..835D, 2011MNRAS.410.2481R, 2011MNRAS.418..467R, 2012MNRAS.419.1163B, 2013MNRAS.435...18B,2013MNRAS.430..370R,doi:10.1137/040614359,2008MNRAS.383..539M}. 

The needlet decomposition allows the ILC weights to vary smoothly on large angular scales and rapidly on small angular scales. The needlet windows in harmonic space, $h^{j}_{\ell}$, are defined as follows 
\begin{eqnarray} 
h^{j}_{\ell} = \left\{
\begin{array}{rl} 
\cos\left[\left(\frac{\ell^{j}_{\rm peak}-\ell}{\ell^{j}_{\rm peak}-\ell^{j}_{\rm min}}\right)
\frac{\pi}{2}\right]& \text{for } \ell^{j}_{\rm min} \le \ell < \ell^{j}_{\rm peak}\,,\\ 
\\
1\hspace{0.5in} & \text{for } \ell = \ell_{\rm peak},\\
\\
\cos\left[\left(\frac{\ell-\ell^{j}_{\rm peak}}{\ell^{j}_{\rm max}-\ell^{j}_{\rm peak}}\right)
\frac{\pi}{2}\right]& \text{for } \ell^{j}_{\rm peak} < \ell \le \ell^{j}_{\rm max} 
\end{array} \right. \,. 
\end{eqnarray}
In terms of $h_{\ell}^{j}$, the spherical needlets are defined as
\begin{eqnarray}
\Psi_{j k}(\hat n)=\sqrt{\lambda_{jk}}\sum _{\ell=\ell^{j}_{\rm min}}^{\ell^{j}_{\max }}\sum_{m=-\ell}^{\ell} 
h_{\ell}^{j}\,Y_{\ell m}^{*}(\hat n)\,Y_{\ell m}(\hat\xi_{jk})\,,
\end{eqnarray}
where the $\{\xi_{jk}\}$ denote a set of cubature points on the sphere for scale $j$. In practice, we identify these points with the pixel centers of the \texttt{HEALPix} pixelization scheme \citep{2005ApJ...622..759G}. Each index $k$ corresponds to a particular \texttt{HEALPix} pixel, at a resolution parameter $N_{\mathrm{side}}(j)$ specific to that scale $j$. The cubature weights $\lambda_{jk}$ are inversely proportional to the number $N_{j}$ of pixels used for the needlet decomposition, i.e., $\lambda_{jk}=\frac{\displaystyle{4\pi}}{\displaystyle{N_{j}}}$. 
Given a set of needlet functions, any sky map of a spin-$0$ field $X(\hat n)$ (such as the CMB temperature anisotropy or the $E$ and $B$ modes)
can be expressed as
\begin{eqnarray}
X(\hat n) =\sum _{\ell=0}^{\ell_{\max }}\sum_{m=-\ell}^{\ell}X_{\ell m}Y_{\ell m}(\hat n)=\sum _{j}^{}\sum _{k}^{}\beta^{X}_{j k}\Psi_{j k}(\hat n),
\end{eqnarray}
where the needlet coefficients, $\beta^{X}_{j k}$, of the sky map are denoted as
\begin{eqnarray}
\beta^{X}_{j k}=\left <X,\Psi_{j k}\right>=\sqrt{\lambda_{j k}} \sum _{\ell=0}^{\ell_{\max
  }}\sum_{m=-\ell}^{\ell} h_\ell^j\,X_{\ell m}\,\,Y_{\ell m}(\xi _{j k})\,. 
\end{eqnarray}
For each scale $j$, the \NILC\ filter has compact support between the multipoles $\ell^{j}_{\rm min}$ and $\ell^{j}_{\rm max}$ with a peak at $\ell^{j}_{\rm peak}$. The values of $\ell^{j}_{\rm min}$, $\ell^{j}_{\rm peak}$ and $\ell^{j}_{\rm max}$ for different needlet bands used in the analysis are listed in table~\ref{tab:needlet-bands}. The needlet coefficients, $\beta^{X}_{jk}$, are computed on the \texttt{HEALPix} grid points, $\xi_{j k}$, with a resolution parameter, $N_{\rm side}$, equal to the smallest power of $2$ larger than $\ell^{j}_{\rm max}/2$.

\begin{table}
\caption{List of needlet bands used in the present
  analysis.}  \centering \begin{tabular}{c c c c c} \hline\hline
  Band index & $\ell_{\rm min}$ & $\ell_{\rm peak}$ & $\ell_{\rm max}$ & $N_{\rm side}$ \\ [1ex]
  \hline 1 & 0 & 0 & 50 & 32\\ 2 & 0 & 50 & 100 & 64 \\ 3 & 50 & 100 &
  200 & 128 \\ 4 & 100 & 200 & 300 & 128 \\ 5 & 200 & 300 & 400 & 256
  \\ 6 & 300 & 400 & 500 & 256 \\ 7 & 400 & 500 & 600 & 512 \\ 8 & 500
  & 600 & 700 & 512 \\ 9 & 600 & 700 & 800 & 512 \\ 10 & 700
  & 800 & 900 & 512 \\ 11 & 800 & 900 & 1000 & 512\\ [1ex] \hline
\end{tabular} 
\label{tab:needlet-bands} 
\end{table}

\bibliographystyle{ptephy}
\bibliography{references}

\end{document}